\documentclass[12pt,preprint]{aastex}

\usepackage{emulateapj5}
\usepackage{graphicx}
\usepackage{fancyheadings}
\usepackage{color}
\usepackage{natbib}

\newcommand{\aox}{\ifmmode \alpha_{ox}\else$\alpha_{ox}$\fi}
\newcommand{\mone}{\ifmmode ^{-1}\else$^{-1}$\fi}
\newcommand{\mtwo}{\ifmmode ^{-2}\else$^{-2}$\fi}
\newcommand{\degs}{\ifmmode ^{\circ}\else$^{\circ}$\fi}
\newcommand{\mv}{\ifmmode {m_{V}}\else${m_{V}}$\fi}

\newcommand{\gae}{\mathrel{\raise .4ex\hbox{\rlap{$>$}\lower 1.2ex\hbox{$\sim$}
} }}

\shorttitle{X-ray cluster at z=1.063}
\begin{document}

\title{X-ray Cluster Associated with the z=1.063 CSS Quasar 3C~186:\\ The Jet is NOT Frustrated.}

\author{Aneta Siemiginowska\altaffilmark{1}, C.~C. Cheung\altaffilmark{2,3}, Stephanie LaMassa\altaffilmark{1}, D.~J. Burke\altaffilmark{1}, Thomas L. Aldcroft\altaffilmark{1}, Jill Bechtold\altaffilmark{4}, Martin Elvis\altaffilmark{1}, D.M. Worrall\altaffilmark{5}}

\altaffiltext{1}{Harvard-Smithsonian Center for Astrophysics,
60 Garden St., Cambridge, MA 02138; asiemiginowska@cfa.harvard.edu}

\altaffiltext{2}{MIT Kavli Institute for Astrophysics \& SpaceResearch
77 Massachusetts Ave.,Cambridge, MA 02139; ccheung@space.mit.edu}

\altaffiltext{3}{Jansky Postdoctoral Fellow; National Radio Astronomy Observatory}

\altaffiltext{4}{Steward Observatory, University of Arizona, Tucson, AZ}

\altaffiltext{5}{Department of Physics, University of Bristol, Tyndall Ave., Bristol, UK}


\begin{abstract}

We report the {\it Chandra} discovery of an X-ray cluster at redshift
$z = 1.063$ associated with the Compact Steep Spectrum radio loud
quasar 3C~186 (Q0740+380). Diffuse X-ray emission is detected out to
$\sim 120$~kpc from the quasar and contains 741$\pm40$ net counts. The
X-ray spectrum of the extended emission shows strong Fe-line
emission (EW=412eV) at the quasar redshift and confirms the thermal
nature of this diffuse component. We measure a cluster temperature of
5.2$^{+1.2}_{-0.9}$~keV and an X-ray luminosity L$_{(0.5-2\rm keV)}
\sim 6 \times 10^{44}$~erg~sec$^{-1}$, which are in agreement with the
luminosity-temperature relation for high-redshift clusters. This is
the first detection of a bright X-ray cluster around a luminous
(L$_{\rm bol}\sim$10$^{47}$~erg~sec$^{-1}$) CSS quasar at high redshift
and only the fifth $z>1$ X-ray cluster detected. We find that the CSS
radio source is highly overpressured with respect to the thermal
cluster medium by about 3 orders of magnitude. This provides direct
observational evidence that the radio source is not thermally confined
as posited in the ``frustrated'' scenario for CSS sources. Instead, it
appears that the radio source may be young and we are observing it at
an early stage of its evolution. In that case the radio source
could supply the energy into the cluster and potentially prevent its
cooling.

\end{abstract}
\keywords{quasars: individual (3C~186) - X-rays: galaxies: clusters}

\section{Introduction}

Powerful radio sources, that are compact on galaxy scales, the
Giga-Hertz Peaked Spectrum (GPS) and Compact Steep Spectrum (CSS)
sources, comprise a significant fraction of the bright radio source
population (10-20$\%$, O'Dea 1998). They are strong candidates for
being the progenitors of large-scale radio sources (e.g. Fanti et
al. 1995, O'Dea \& Baum 1997, O'Dea 1998), but this connection has not
been firmly established. Their radio morphologies show compact
emission on arcsec (VLA resolution) scales while on milliarcsec scales
(VLBI) the sources look remarkably like scaled down large radio
galaxies, where the entire radio structure (1-10~kpc) is enclosed
within the host galaxy. Since the first GPS samples have been
constructed there has been a clear controversy regarding their nature
(see O'Dea 1998 and references therein).  In the {\it evolution} model
the source size and the characteristic spectral break at GHz radio
frequencies could be an indication of young age, while in the other
model the radio jet could be {\it frustrated} by a dense confining
medium. Recent observations (e.g. measured expansion timescales of
$<1000$ years, Owsianik et al. 1998, Polatidis \& Conway 2003) give
more weight to the {\it evolution} model (Readhead et al. 1996,
Snellen et al. 2000, Alexander 2000), although there has been no
definite observational evidence to rule out either of the models and
both interpretations are still viable.

Here, we report the {\it Chandra} discovery of extended X-ray emission
associated with the compact steep spectrum (CSS) quasar, 3C~186
(Q0740+380, z=1.063). The {\it Chandra} spectrum of the diffuse
emission contains 741$\pm40$ counts and a strong (EW$\sim$412eV) iron
emission line at the quasar redshift characteristic of thermal
emission. This is the first observation of thermal emission associated
with a CSS quasar at high redshift and gives us a rare opportunity to
study interactions between an expanding CSS radio source and the
cluster medium. As we show in Section~6.2 that the pressure of the hot
cluster gas is too low, by 2-3 orders of magnitudes, to confine the
radio source.  This is direct observational evidence that the radio
source is not thermally confined, and so instead is presumably young
so that we are observing it at an early stage of its evolution.

Over the last decade attempts have been made to find X-ray clusters
associated with radio-loud sources at high redshift (e.g. O'Dea 2000,
Siemiginowska et al 2003 for studies related to GPS and CSS). If the
clusters are found around large number of radio-loud sources then this
could be used to place constraints on structure-formations models at
large redshift.  Bremer, Fabian \& Crawford (1997) describe the model
for an onset of a powerful radio source in the center of a cooling
flow cluster. Such a cluster could also confine a compact radio
source.  The limited capabilities of the available X-ray telescopes
allowed only for a few detections of extended X-ray emission around
radio sources at redshifts $z>0.3$ (Hardcastle
\& Worrall 1999, Crawford \& Fabian 2003, Worrall et al. 2001). 
High dynamic range observations are required to detect faint diffuse
emission in the vicinity of a bright powerful source.  The {\it
Chandra} X-ray Observatory can resolve spatially distinct X-ray
emission components in the vicinity of a strong X-ray source with
$\sim$1~arcsec resolution and a high dynamic range, as evidenced, for
example, by the discovery of many resolved quasar X-ray jets
(e.g. Schwartz et al. 2000, Siemiginowska et al. 2002, Sambruna et
al. 2004, Marshall et al. 2005).

The large-scale X-ray emission observed in radio sources can result
from several processes, confusing detections of X-ray clusters at
$z>1$ (e.g. Celotti \& Fabian 2004). Radio synchrotron emission implies
the presence of a population of relativistic particles that produce
high-energy emission via inverse Compton scattering of the cosmic
microwave background (CMB) photons.  The energy density of the CMB
increases with redshift as $(1+z)^4$, so the surface brightness of the
inverse Compton emission is approximately constant with redshift. This
is in contrast with thermal cluster emission where the surface
brightness drops with redshift.  X-ray spectral and spatial
information are key to identifying different emission components and
measuring diffuse thermal X-ray emission.

At the highest redshifts Carilli et al. (2002) describe extended
emission possibly associated with thermal emission from shock-heated
gas within $\sim 150$~kpc of the $z=2.156$ radio galaxy PKS~1138-262.
Fabian et al. (2003) report $\sim$100~kpc-scale emission around the
nucleus of the redshift $z=$1.786 radio galaxy 3C~294 and give several
possible explanations for the origin of this emission. In both cases
the poor quality of the X-ray spectrum does not allow confirmation of
the thermal nature of the diffuse X-ray emission. At redshifts
$0.5<z<1.0$ thermal confirmation has been possible in some sources,
e.g. 3C~220.1 (Worrall et al. 2001)

Several deep images of nearby X-ray clusters (Fabian et al. 2003,
Forman et al. 2003, Nulsen et al. 2004) obtained recently with {\it
Chandra} provide evidence that intermittent AGN outbursts with an
average power of $\sim10^{45}$~erg~sec$^{-1}$ supply energy into the
cluster preventing its cooling (McNamara et al. 2005).  These deeply
imaged clusters contain only relatively low power AGN ($\sim
10^{40}-10^{43}$ erg~sec$^{-1}$) that are interpreted as having been
active in the past in order to drive the observed cluster
morphology. In contrast to these low redshift clusters, the 3C~186
cluster is observed during the quasar's active phase while heating the
cluster medium.

3C~186 is a very luminous quasar (L$_{bol} \sim 10^{47}$
erg~sec$^{-1}$). It has a strong big blue bump in the optical-UV band
and broad optical emission lines (Netzer et al. 1997, Simpson \&
Rawlings 2000, Kuraszkiewicz et al. 2002, Evans \& Koratkar 2004). It
is therefore a typical quasar except for its radio properties. The
radio morphology shows two components separated by 2$\arcsec$ and a
jet connecting the core and NW component (Cawthorne et al.,
1986). Murgia et al. (1999) estimated the age of the CSS source to be
of the order of $\sim 10^5$~years based on the spectral age of the
radio source. Our observation provides X-ray morphology and spectral
information for the quasar and an associated X-ray cluster. The quasar
core is so bright in X-rays that any X-ray emission associated with
the radio components is not spatially resolved.  The diffuse X-ray
cluster emission is detected beyond the quasar core and the CSS
source.

We describe the {\it Chandra} observation, our analysis techniques and
the results for the extended component in Sec.2. Section 3 and 4
present the archival radio and optical data.  The quasar X-ray
spectrum is presented in Sec.5. Section 6 contains the discussion.

Throughout this paper we use the cosmological parameters based on the
WMAP measurements (Spergel et al. 2003):
H$_0=$71~km~sec$^{-1}$~Mpc$^{-1}$, $\Omega_M = 0.27$, and $\Omega_{\rm
vac} = 0.73$. At $z= 1.063$, 1\arcsec\ corresponds to $\sim$8.2~kpc.

\section{Chandra Observations}

3C~186\ was observed for $\sim 38$~ksec with the {\it Chandra}
Advanced CCD Imaging Spectrometer (ACIS-S, Weisskopf et al. 2002) on
2002 May 16 (ObsID 3098).  The source was located $\sim 35\arcsec$
from the default aim-point position (to avoid node boundaries) on the
ACIS-S backside illuminated chip S3 (Proposer's Observatory Guide
(POG)\footnote{http://asc.harvard.edu/proposer/POG/index.html}). The
1/8 subarray CCD readout mode of one CCD only was used resulting in
0.441~sec frame readout time.  The observation was made in VFAINT mode
with the standard 5x5 pixel island used to assign the event grades by
the pipeline. This mode allows for a more efficient way of determining
the background events and cleaning the background, especially at the
higher energies. After standard filtering the effective exposure time
for this observation was 34,398~sec. Given the ACIS-S count rate of
0.025~counts~s$^{-1}$~frame$^{-1}$ the pileup fraction was low $<2\%$
(see PIMMS\footnote{http://asc.harvard.edu/toolkit/pimms.jsp}).

\begin{figure}
\epsscale{0.85}
\plotone{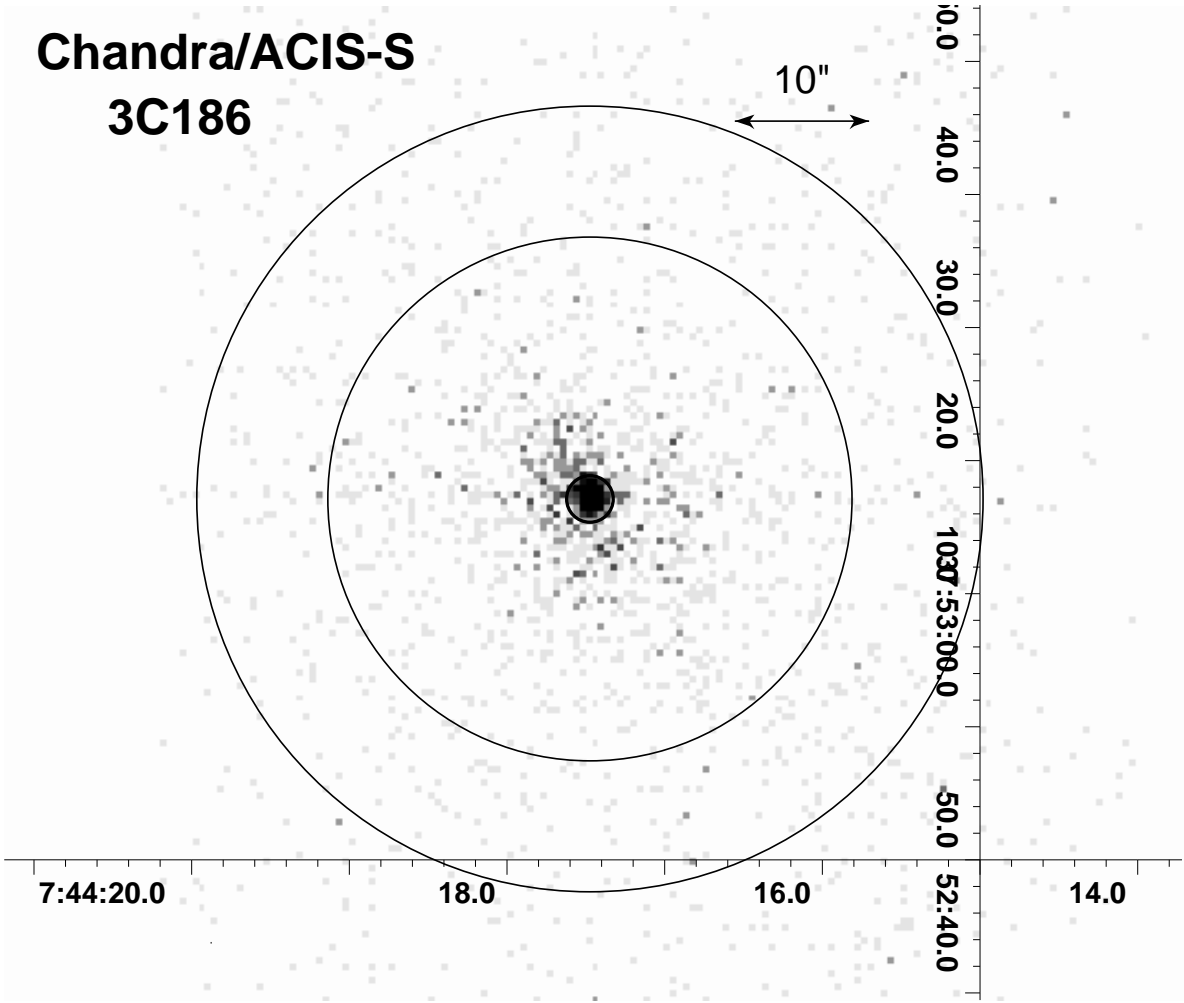}
\caption{ {\it Chandra} ACIS-S X-ray (0.3-7~keV) image of 3C~186.
The pixel size is the standard ACIS-S pixel of 0.492\arcsec. The
quasar source region assumed for extracting the spectrum is a
1.75\arcsec radius circle in the center of the field. The outer
annulus with 20\arcsec\ and 30\arcsec\ radii illustrates the
background region.  The arrow in the upper right corner indicates the
10\arcsec scale. North is up and East is left. Coordinates are J2000.}
\label{fig:acis1}
\end{figure}

Figure~\ref{fig:acis1} shows the {\it Chandra} ACIS-S image overlayed
with the quasar core and background regions. Figure~\ref{fig:smooth}
shows the ACIS-S image of 3C~186 adaptively smoothed with the CIAO
tool CSMOOTH. The X-ray emission is more extended than the X-ray
emission of a typical radio-quiet quasar observed with {\it Chandra}.
The diffuse emission extends up to $\sim 120$~kpc in radius.
Figure~\ref{fig:comparisons} show the smoothed images in soft
(0.5-2~keV) and hard (2-7~keV) energy bands. Only the NE quadrant is
visible in the hard band, while all of the extended emission is
visible in the soft. Below we describe in detail the imaging and
spectral analysis of this emission.

\subsection{Imaging Analysis}

The X-ray data analysis was performed in CIAO
3.2\footnote{http://cxc.harvard.edu/ciao/} with the calibration files
from the CALDB 3.0 data base. Note that the ACIS-S contamination file
{\tt acisD1999-08-13contamN0003.fits} was included in our analysis;
this accounts for the temporal, but not spatial, variation of the
contamination layer on the optical-blocking filter of
ACIS\footnote{http://cxc.harvard.edu/ciao/why/acisqedeg.html}.  Since
the quasar is observed close to the aim-point, which is where the
spatially-invariant contamination model was calibrated, the results
will not change if the data were re-analyzed using the
spatially-dependent contamination model. We used {\it Sherpa} (Freeman
et al. 2001) for all spectral and image modeling and fitting.

We ran {\tt acis\_process\_events} to remove pixel randomization and
to obtain the highest resolution image data. The X-ray position of the
quasar (J2000: 07 44 17.47 +37 53 17.11) agrees with the radio
position (Li \& Jin 1996) to better than 0.1$\arcsec$, (which is
smaller than {\em{Chandra}}'s 90$\%$ pointing accuracy of 0.6~arcsec,
Weisskopf et al. 2003), so we have high confidence in the source
identification.

To determine the size of the extended emission we ran a ray trace
using CHaRT\footnote{http://cxc.harvard.edu/chart/} and then
MARX\footnote{http://space.mit.edu/CXC/MARX/} to create a high S/N
simulation of a point source.  We modeled the quasar core as a point
source with the energy spectrum given by the fitting described in
Section~5. We then extracted a radial profile from both the {\it
Chandra} data and the simulated point source image assuming annuli
separated by 1 arcsec and centered on the quasar.  For the simulation
we added 7$\%$ errors to account for uncertainty in the raytrace model
(Schwartz et al 2000a, Jerius et al 2004). The PSF was normalized to
match the peak surface brightness of the core. The resultant profiles
are shown in Fig~\ref{fig:sim1} and clearly illustrate that the
observed emission (empty squares) is highly inconsistent with a point
source (solid line and empty triangles).

We apply a $\beta$-model to the one-dimensional surface brightness
profile beyond $\sim 3\arcsec$ radius to estimate a core radius and
$\beta$ parameter for the diffuse emission. The best fit model is
represented in Figure~\ref{fig:sim2} with
$\beta$=0.64$^{+0.11}_{-0.07}$ and a core radius
$r_c=5.8$\arcsec$^{+2.1}_{-1.7}$ which corresponds to
47$_{-14}^{+13}$~kpc.  

Since the core radius found in the one-dimensional analysis is small,
a two-dimensional fit was made to see what influence the quasar
emission and any non-sphericity of the cluster emission has on the
$\beta$-model parameters. A model consisting of the ChaRT-generated
PSF and a two-dimensional $\beta$ model was fitted to the data, using
the default pixel size of 0.492\arcsec. The Cash statistic (Cash 1979)
was used, since the number of counts per pixel was low outside the
core, and the cluster center, ellipticity, and position angle were
allowed to vary as well as the normalization, core radius, and $\beta$
parameter. The best-fit cluster location differs from that of the
quasar by 0.2\arcsec, which is within the one-sigma error circle of
the position (0.3\arcsec). The best-fit model is elliptical, with an
ellipticity of $0.24_{-0.07}^{+0.06}$ and position angle of
$47\pm{10}$~degrees, but the core radius and $\beta$ values are
similar to the one-dimensional results, with $r_c =
5.5_{-1.2}^{+1.5}$\arcsec\ (45$_{-10}^{+12}$~kpc) and $\beta =
0.58_{-0.05}^{+0.06}$.

\subsection{Spectral Analysis of the Extended emission} 

We extracted the energy spectrum of the extended emission from an
annulus of radii 2.7\arcsec\ and 15\arcsec\ centered on the
quasar. There are 1189$\pm 34$ total counts and 741.4$\pm 40.4$ net
source counts in this region in the full {\it Chandra} energy
band. The background spectrum was taken from the annulus of radii
20\arcsec\ and 30\arcsec\,. Because the background increases at low
and high energies we modeled only the spectrum within a 0.3-7~keV
energy range.  The total number of counts in this energy range was
876$\pm 31$ with 691.0$\pm 32.5$ net counts. In all spectral modeling
we simultaneously fit background and source data applying $\chi^2$
statistics with the weighting described by Primini et al. (1994) as
implemented in {\it Sherpa}. All errors quoted below are $90\%$ errors
for a single parameter calculated with the {\tt projection} routine in
{\it Sherpa}. Table 1 lists the applied models and the best fit
values.

We first fitted the spectrum of the diffuse emission with an absorbed
power-law model assuming the equivalent column of hydrogen in the
Galaxy of 5.68$\times 10^{20}$~atoms~cm$^{-2}$ (COLDEN\footnote{Stark
et al 1992}). We then tested for excess absorption.  The fitted
absorbing column is in agreement with the Galactic value (see Table
1). Modeling the data with a power law indicated an excess at the
Fe-line energy characteristic of thermal emission at the quasar
redshift. We added a gaussian line to the model and obtained the line
equivalent width of EW=412~eV at the observed energy
E$_{obs}$=3.18${\pm 0.07}$~keV (90$\%$ errors) corresponding to the
rest-frame energy of 6.56$\pm 0.14$~keV. Note that this line energy
indicated that the line is emitted by the hot ionized plasma.

We next applied the {\tt RAYMOND} and {\tt MEKAL} XSPEC plasma models
in {\it Sherpa} and obtained gas temperatures 4.4-5.2~keV for two
choices of abundance, solar and 0.3 solar, using the abundances of
Anders \& Grevesse (1989). Figure~\ref{fig:raymond} shows the best-fit
thermal model together with the residuals. The observed unabsorbed
flux assuming these models is of order 6.2$\pm0.3 \times
10^{-14}$erg~sec$^{-1}$~cm$^{-2}$ for the 0.5-2~keV energy range and
5$\pm0.7 \times 10^{-14}$erg~sec$^{-1}$~cm$^{-2}$ for the 2-10~keV
energy range (the flux errors are based on 5$\%$ and 14$\%$ counts
uncertainties in each band respectively).

The 0.5-2~keV rest frame X-ray luminosity of the cluster emission is
equal to $\sim 3 \times 10^{44}$~erg~sec$^{-1}$ (K-corrected).  This
is only the luminosity calculated based on the emission in the adopted
annulus extending to about 122~kpc. This is about 52$\%$ of the total
cluster luminosity as estimated from the radial profile fitting
described in the previous section. Thus the total cluster luminosity
is approximately $\sim 6 \times 10^{44}$~erg~sec$^{-1}$.

To investigate a possible non-thermal contribution to the spectrum due
to inverse Compton scattering of CMB photons we fixed the {\tt MEKAL}
model at the best value and added a power law component with $\Gamma=1.7$
to the model. The 3$\sigma$ upper limit for the contribution of this
component at 1~keV is then 3.4$\times 10^{-6}$
photons~cm$^{-2}$~sec$^{-1}$, i.e. less than $12\%$ of the extended
emission can be non-thermal.

To investigate possible temperature variations in the radial direction
we extracted the spectra from two annuli: the inner one spans
2.7\arcsec-7.8\arcsec\ and the outer one spans
7.8\arcsec-15\arcsec. Table 2
summarizes the results. The hardness radio indicates that there are
more soft counts in the inner region than in the outer one. The {\tt
RAYMOND} model fit to these spectra indicate a slight temperature
decrease, by $\Delta \rm kT \sim 0.3$~keV, towards the inner radii.
However the errorbars are larger than any apparent deviations. The
available data do not allow for fitting of cooling flow models (see
Sec.6.1 for cooling-flow discussion).

To further investigate possible temperature variations of the extended
emission in the azimuthal direction we divided the 2.7\arcsec\ to
15\arcsec\ annulus into four sectors and extracted counts and spectra
from these sectors. The sectors, illustrated in Fig.\ref{fig:panda}a,
were chosen to follow the non-symmetrical shape of the extended
emission as apparent in the smoothed image (Fig.\ref{fig:smooth} and
Fig.\ref{fig:comparison}). We calculated the surface brightness
profile in each sector which confirmed that the emission in the NE-SW
direction (sectors 1 and 3) is stronger (by a factor of 1.25-1.5) than
in NW-SE direction (sectors 4 and 2). The surface brightness values
are indicated in Figure~\ref{fig:panda}b.

The X-ray emission properties of the sectors are presented in
Table 2.
The hardness ratios indicate possible spectral differences between
sectors. We fit the {\it Chandra} spectrum of each sector with a
thermal emission model. There is a slight temperature variation (with
a minimum of 3.7~keV and a maximum of 4.9~keV) between sectors in the
best-fit values, however the error bars are approximately 1.5-2~keV,
so that the fitted variations are all within 90$\%$ errors.

In summary, the diffuse emission is non-symmetric and elongated in the
NE-SW direction with the harder emission towards NE (see Fig.3).    Note
that this structure is orthogonal to the radio emission described in
the next section.

\section{Radio Observations}

We searched for radio emission corresponding to the diffuse X-ray
emission revealed in the {\it Chandra} image by reanalyzing VLA 1.5
GHz data published by van Breugel et al.. (1992).  The
multi-configuration (A and B) data were obtained in 1987 and amounted
to a total integration time of about 80 minutes split almost equally
between the two configurations. At the resolution of our image
(Figure~\ref{fig:radio}a), the radio source is dominated by a 1.8''
double source aligned at a position angle of about --37 deg.  Our
self-calibrated dataset does not show extended radio emission on the
scale of the X-ray emission detected above a 3$\sigma$ rms noise of
0.75 mJy/beam in the naturally weighted image; the off-source
($>$15'') rms achieved in the image is about a factor of two smaller.

We also analyzed a 410 second VLA 15 GHz A configuration dataset
of 3C~186 obtained on 13 Dec 1992 (program AL280) in order to examine
the radio jet more clearly.  The image shows similar features to
previous high resolution maps (Cawthorne et al.. 1986; Spencer et
al. 1991), i.e., a one-sided jet to the northwest connecting a
flat-spectrum core to a diffuse radio lobe, and a bright radio lobe in
the southeast direction (see Fig.~\ref{fig:radio}b).


\section{Hubble Space Telescope Data and Optical Emission}

A cosmic-ray rejected WFPC2
Associations\footnote{http://archive.stsci.edu/hst/manual} image with
a total integration time of 8,000 seconds was downloaded from the HST
archive. The F675W image, obtained in Program 6491, was centered on
one of the lower resolution wide field (WF) chips.  We modeled the
central source with elliptical isophotes utilizing the ELLIPSE task in
the
STSDAS\footnote{http://www.stsci.edu/resources/software\_hardware/stsdas}
package and subtracted it from the image in order to show the nearest
lying objects more clearly.  The resultant image (Fig.\ref{fig:hst})
shows several sources within the X-ray cluster emission indicated by a
15$\arcsec$ circle. Unfortunately only a single band image of 3C~186
was taken with the WFPC2 camera and we cannot identify the colors of
these sources.

\section{Quasar X-ray emission}

The quasar core emission dominates the overall X-ray emission in the
vicinity of 3C~186. We define the quasar emission region as a circle
with 1.75\arcsec\ radius and assume background emission from an
annulus with radii 20\arcsec and 30\arcsec\ (see
Fig.\ref{fig:acis1}). Based on the PSF modeling we estimate that
$\sim$98\% of the point-source counts are included in this source
region. The extracted quasar spectrum contains 1968.7$\pm 44.3$ net
counts (1905.4 net counts in energy range between 0.3-7~keV). We model
the spectrum with an absorbed power law.  The best-fit power-law model
has a photon index $\Gamma = 2.01{\pm 0.07}$ and a 2-10~keV flux of
1.7$\times 10^{-13}$~ergs~cm$^{-2}$~sec$^{-1}$ which corresponds to
a quasar X-ray luminosity L$_X$(2-10~keV) = 1.2$\times
10^{45}$erg~sec$^{-1}$ and L$_X$(0.5-2~keV)=1.1$\times
10^{45}$erg~sec$^{-1}$ (unabsorbed and K-corrected luminosity). We do
not detect any significant neutral absorbing column intrinsic to the
quasar with the 3$\sigma$ upper limit to the equivalent column of
Hydrogen of $<9.0\times 10^{20}$atoms~cm$^{-2}$.

The power-law fit to the data leaves some residuals at $\sim 3$~keV
indicating a possible emission line at this energy. We added the
Gaussian line component to the model and obtained the best fit
location for the narrow emission line (FWHM$ <0.23$~keV) at
E$_{obs}$=3.07$^{+0.06}_{-0.11}$~keV corresponding to
E$_{rest}$=6.33$\pm0.06$~keV. The line equivalent width is equal to
EW=162~eV and it can be identified with Fe-K${\alpha}$ emission.
The best-fit model and residuals are shown in
Fig.\ref{fig:qsofit}. (Note that the residuals between 1 and 2 keV are
due to calibration uncertainties.)  We can estimate the flux
contribution to this line from the extended thermal emission by
extrapolating the radial profile of the extended emission into the
central circular region assumed for the quasar emission. The
contribution from the thermal cluster emission to the quasar spectrum
is of order 10$\%$. Given the line flux in both components we estimate
that only about $\sim$5$\%$ of the line emission can come from the
thermal cluster emission, thus the line is dominated by the nuclear
emission. Note that the energy, E$_{rest}$=6.33$\pm0.06$~keV, of the
detected Fe-line in the quasar spectrum is in agreement with being
emitted by the neutral medium, while the energy of the Fe-line emitted
by the cluster gas, E$_{rest}$=6.56$\pm0.14$~keV, indicates an emission
from ionized gas.

\section{Discussion}

\subsection{X-ray Cluster}

Our {\it Chandra} observation reveals X-ray cluster emission at the
redshift of the quasar 3C~186. The X-ray properties of the cluster are
summarized in Table 3.  We compare the cluster temperature and its
luminosity with results for the other clusters at high redshift using
the {\tt MEKAL} model with the abundance set to 0.3 as in Vikhlinin et
al.(2002).  The cluster temperature of $\sim5.2^{+1.2}_{-0.9}$~keV and
the total X-ray luminosity of L$_X(0.5-2~\rm keV)$ $\sim 6 \times
10^{44}$~erg~sec$^{-1}$ agree with the temperature-luminosity relation
typically observed in high redshift ($z>0.7$) clusters (e.g. Vikhlinin
et al.  2002, Lumb et al. 2004).

We can estimate physical properties of the cluster using standard
formulae (Donahue et al. 2003; Worrall \& Birkinshaw,
2004). Approximating the X-ray diffuse emission as spherical, we
calculate the cluster central electron density ($n_0$) to be
approximately 0.044$\pm0.006$~cm$^{-3}$ (errors are only due to the
uncertainty in normalization) for the best-fit 1D beta model
parameters: $\beta$=0.64, a core radius of $\theta_c=5.8\arcsec$
(Sec.2.1), a gas temperature of 5~keV and the spectral normalization of
3.5$({\pm 0.4})\times$10$^{-4}$ based on the {\tt MEKAL} model fit to
the spectrum of the diffuse emission (assuming the annulus between
2.7\arcsec and 15\arcsec radii).  These parameters imply that the mass
of the gas enclosed within 1~Mpc radius of the isothermal sphere is
$\sim$2.2$\times 10^{13}$M$_{\odot}$. The total gravitational mass
enclosed within 1~Mpc is $\sim$2.6$\times 10^{14}$M$_{\odot}$ assuming
an isothermal-sphere model. The cluster gravitational mass is
comparable to the mass of the other high redshift clusters recently
measured with {\it Chandra} and XMM-{\it Newton} (Donahue et al. 2003,
Worrall \& Birkinshaw 2003, Vikhlinin et al 2002).  The gas fraction,
i.e. $\sim 10\%$, broadly agrees with the gas fraction usually found
in high redshift (z$>$0.7) clusters.  

We found a relatively small core radius, $\sim 50$~kpc, for the 3C~186
cluster while most redshift $z>0.7$ clusters have typical core radii
larger than $100$~kpc. If gas of the same X-ray luminosity as found
within a radius of 123~kpc of 3C~186 were distributed with a beta
model of more typical core radius (150~kpc), the gas mass contained
within a radius of 1~Mpc would increase by a factor of about 2.

Is there a cooling flow in this cluster? 
We estimate the gas mass enclosed within the core radius of $\sim
50$~kpc to be of the order of 2.2$\times 10^{11}$M$_{\odot}$. Given
the gas central density and the temperature of 5~keV the cooling time
for this core is $\sim 1.6
\times 10^9$~years and without a heat source there would be a cooling
flow with the cooling-flow rate $\sim 50$ M$_{\odot}$~yr$^{-1}$
(Fabian \& Nulsen 1977, Fabian 1994).

Our image analysis indicates that the cluster X-ray emission follows
an elliptical distribution (see Section 2.1). The angle between the
major axes of the two-dimensional ellipse and the radio jet axes is
$\sim 84_{-6}^{+7}$~degrees. This type of X-ray vs. radio morphology
is often seen in lower redshift clusters and it is interpreted as due
to interactions between the radio plasma and the cluster medium (as
for example in Hydra cluster in Nulsen et al 2002). The current radio
data do not show any radio emission on scales similar to the observed
X-ray cluster emission. However, the X-ray morphology suggests that
there could be old plasma there possibly associated with previous
activity of the quasar.  The detection of a relic at low radio
frequency observation would allow studies of the evolution timescales
and the feedback between the quasar activity and the cluster medium.

\subsection{Optical Environment of 3C~186}

Optical observations of the quasar 3C~186 indicated that the source is
located in a rich galaxy environment. S\'anchez and Gonz\'alez-Serrano
(2002) studied an over-density and clustering of galaxies in the
optical field of several high redshift radio sources. They include a
K-band image of the 3C~186 field and indicate that the surface density
of the galaxies in a possible cluster peaks to the NE about
50$\arcsec$ from the quasar. While the reported location of the peak
density is located just outside the {\it Chandra} FOV we do not detect
any significant X-ray emission towards the optical peak away from the
quasar. In fact the quasar seems to be centrally located with respect
to the X-ray diffuse emission.  

Recent optical study of the cluster environments of radio-loud quasars
at $0.6<$z$<1.1$ by Barr et al (2003) shows that these quasars do not
reside in the center of the galaxy distributions. Because the X-ray
emitting gas traces the cluster's gravitational potential, the X-ray
observations can confirm whether the quasar is located in the center
of the potential well of the cluster.  The {\it Chandra} observations
3C~186 indicate that the quasar is located at the center of the X-ray
emission and offset from the center of the galaxy distribution. 

\subsection{Lack of Confinement of the CSS source}

The diffuse emission on a $\sim$120~kpc scale discovered in this {\it
Chandra} observation indicates the presence of a hot Intercluster
medium (ICM) surrounding this powerful CSS quasar.  The X-ray spectrum
of the cluster is dominated by a thermal component with a strong
Fe-line at the quasar redshift.  In one scenario the small size of CSS
radio sources is associated with a dense environment that prevents the
expansion of the source and confines the radio lobes to the size of a host
galaxy (Wilkinson et al. 1981, van Breugel et al. 1984, O'Dea et
al. 1991).  Is this cluster then responsible for confining the CSS
source?

Based on the cluster central density and temperature, we estimate a
central thermal pressure of $\sim 5 \times 10^{-11}$~dyn cm$^{-2}$. If
this pressure is higher than the pressure within the expanding radio
components of the CSS source then the cluster gas may be responsible
for confining the radio source and its small size.

The radio source is dominated by a double at 1.5 GHz, which straddles
a central flat-spectrum radio core as seen in the higher resolution 15
GHz map (Figure~\ref{fig:radio}). Using DIFMAP's MODELFIT (Shephard,
Pearson \& Taylor 1994) routine, the 1.5 GHz data are best-fit with
two 0.59 Jy elliptical Gaussians with dimensions of 0.7'' x 0.3'', and
a separation of 1.8''.  Taking the spectral index to be 1, and
approximating each component as a homogeneous spheroid, we estimate
the minimum pressure in each radio component to be $\sim$10$^{-8}$ dyn
cm$^{-2}$ (see also Murgia et al 1999).  Thus the radio source is
highly overpressured by about 2-3 orders of magnitude with respect to
the thermal cluster medium.

The new X-ray detection gives direct observational evidence that the
radio source is not thermally confined as posited in the "frustrated"
scenario for CSS sources.  Instead, at least in this CSS quasar, it
appears that the radio source may indeed be young (Readhead \& Hewish
1976; Phillips \& Mutel 1982, Carvalho 1985) and that we are observing
an early stage of radio source evolution.

The jet can also interact and be stopped by a clumpy cold medium of
the host galaxy, as described by Carvalho et al 1998, De Young 1991 or
Jeyakumar et al 2005. The absorption due to this cold medium should be
detectable in the quasar X-ray spectrum. We can use the 3$\sigma$
upper limit of 9$\times 10^{20}$ cm$^{-2}$ on the total absorbing
column density intrinsic to the quasar (see Sec.5 and Table 4) to
estimate the total size of the clumpy medium.  The {\it frustrated}
jet models require clouds with densities of 1-30~cm$^{-3}$. The
absorption limit gives the size of the clumpy medium of order
10-100~pc compared with the 16~kpc diameter of the radio source. Any
such region cannot significantly limit the expansion of the radio
source and frustrate the jet (see also discussion in Guainazzi et al
2004).

Our X-ray observation of 3C~186 indicates that the CSS radio source is
not confined, but it is at its early stage of the evolution into a
large scale radio source.

\subsection{The CSS and the Cluster Heating}

The CSS radio components are overpressured with respect to the thermal
cluster gas. Thus the expansion of these components into the cluster
medium could potentially heat the center of this cluster. The energy
dissipated into the cluster by the expanding radio components has been
widely discussed in the context of the low redshift clusters, where
there is evidence for the repetitive outbursts of an AGN. However, the
details of the dissipation process are undecided with ion viscosity
and ``sound'' waves being possible candidates (Fabian et al. 2005). On
the other hand the mechanical energy released during the shock wave
propagation throughout the cluster could be transfered into the
cluster thermal energy at the location of the shock (P.Nulsen private
communication).
 

We can estimate the energy content of the hot cluster gas assuming a
total emitting volume of 2.3$\times 10^{71}$cm$^3$ (contained by an
annulus with 3 and 15$\arcsec$ radii, assuming spherical geometry) and
$kT \sim 5$~keV, to be of the order of ${3\over2} kTnV \sim$ 2$\times
10^{61}$ ergs (where $n$ is the average gas particle density in the
cluster). We can estimate the jet power from the relation between the
radio luminosity and the jet power given by Willott et al (1999,
Eq.(12)).  Using the 151~MHz flux density of 5.9$\times 10^{-24}$
erg~sec$^{-1}$~cm$^{-2}$~Hz$^{-1}$ (Hales et al. 1993) which accounts
for the total radio emission from the jet and hot spots (the core is
already almost completely self-absorbed at 1.7~GHz, Spencer et
al. 1991) the jet kinetic power is of order $L_{jet} \sim
10^{46}$erg~sec$^{-1}$. If the expanding radio source dissipated the
jet's energy, $L_{jet} \sim 10^{46}$erg~sec$^{-1}$, into the cluster's
central 120~kpc region then the heating time would be $\sim$10$^8$
years.  We can also estimate the amount of mechanical work done by the
jet and radio components during the expansion to the current radio
size ($2\arcsec
\times 0.3\arcsec \sim 2.3 \times 10^{66}$cm$^3$) as $pdV \sim 
10^{56}$~ergs.  If the expansion velocity is of the order of
0.1$c$ then the radio source has been expanding for about $5 \times
10^5$~years with an average power of 6$\times
10^{42}$~erg~sec$^{-1}$. The estimated jet power is $\sim 3$ orders of
magnitude higher.


\subsection{X-ray emission of the core}

The evolution of radio-source expansion within host galaxies and
clusters has been considered by Heinz, Reynolds and Begelman (1998).
They simulated interactions between a growing radio source and the
interstellar and intergalactic medium. For the highly supersonic
expansion of the young source a shock forms around the expanding
source and it heats up the medium to X-ray temperatures. As a
result a ``cocoon'' of hot medium surrounds the radio source.
Depending on the density of the medium and the strength of the shock
a source of the size of 16~kpc can emit
$\sim10^{45}$~erg~sec$^{-1}$ in the {\it Chandra} band.

The double CSS radio source and the radio jet remain unresolved in the
{\it Chandra} observation of 3C~186. Thus the ``cocoon'' region is
located within the unresolved X-ray core and thus could be giving a
significant contribution to the total observed X-ray luminosity of
10$^{45}$~erg~s$^{-1}$. The amount of this contribution depends on the
physical parameters of the expanding radio source and the ISM. A weak
Fe-line is the only emission feature present in the otherwise
featureless X-ray spectrum, however the observed spectral photon index
is quite steep for a typical radio loud quasar (Elvis et al 1985,
Bechtold et al 1994). We therefore estimated a possible thermal
contribution to the quasar luminosity using the {\tt RAYMOND} plasma
model fits to the quasar spectrum. We concluded that about $\sim 15\%$
of the 0.5-2~keV luminosity, e.g. 1.5$\times 10^{44}$ erg~sec$^{-1}$
could be due to the thermal emission. Of course there are other
possible contribution to the observed X-ray spectrum from the radio
jets knots and hot spots.  Consistent theoretical modeling of the
expansion and the emission of the jet components in the future may
help in understanding their relative contributions to the X-ray
spectrum. This is necessary in order to disentangle ``true'' quasar
emission related directly to the accretion flow.

The quasar optical-UV (big blue bump) luminosity of 5.7$\times
10^{46}$erg~sec$^{-1}$ (based on measurements in Simpson \& Rawlings,
2000) is dominated by the typical quasar emission related to the
accretion onto a supermassive black hole. We can therefore estimate
the central black hole mass and required accretion rate based on that
luminosity.  Assuming that the quasar is emitting at the Eddington
luminosity the black hole mass should be of the order $\sim 4.5 \times
10^8$M$_{\odot}$. Based on CIV FWHM measurements of Kuraszkiewicz et
al (2002) and the Vestergaard (2002) scaling relationship for the
black hole mass, the estimated mass of the black hole is approximately
a factor of 10 higher, $\sim 3.2 \times 10^9$M$_{\odot}$.  In any case
the accretion rate required by the observed UV luminosity, and
assuming 10$\%$ efficiency of converting the gravitational energy into
radiation, is equal to $\sim 10$~M$_{\odot}$year$^{-1}$. Given the age
of the radio source of $5 \times 10^5$ years, a total of $\sim 5
\times 10^6$M$_{\odot}$ should have been accreted onto the black hole
to support the current ``outburst''.

Recent studies of formation of galaxies and growth of a supermassive
black hole suggest that the quasar activity is a result of a merger
event (e.g. DiMatteo, Springel \& Hernquist 2005). In this model a
short phase of quasar activity is a direct consequence of the
increased fuel supply onto a central black hole. Based on the radio
aging and the current size of 3C~186 its activity started $\sim
10^6$~years ago. The X-ray cluster emission of 3C~186, although
elliptical in shape does not show any signatures of a merger event
responsible for that activity. The cooling time for this cluster is
$\sim 10^9$~years, so if this cluster is forming then the gas flowing
into the center could onset the radio activity (Bremer et
al. 1997). Such a powerful outburst can easily prevent the cooling of
the cluster core. Although, the exact process of transferring the
outburst energy into the cluster gas is unknown, the available
outburst energy exceeds by $>2$ orders of magnitude the luminosity of
the cluster core. The feedback between the jet heating and the cooling
of the cluster can regulate the growth of the central black hole. On
the other hand the short timescale of the current outburst may also
suggest that the intermittent AGN behaviour could be related to the
physics of the accretion flow (Siemiginowska, Czerny \&
Kostyunin 1996, Janiuk et al 2004) instead of the properties of a
large scale environment triggering the quasar outburst. This is the
first cluster observed at the early phase of the quasar radio activity
giving us a potential to study the early stages of interactions
between the cluster and AGN.

\subsection{Old halo: relic}

Is the current 3C~186 outburst the first phase of the quasar activity
in this cluster?  The large mass of the central black hole suggests
that the supermassive black hole must have been formed much earlier
than 10$^5$~years ago. As we show in the previous section only a small
fraction of the mass has been accreted during the current active
phase.

One way to accommodate the observed statistics of source sizes in the
radio source evolution is to invoke intermittent activity (Reynolds \&
Begelman 1997) with average timescales of $\sim$10$^{4}$ -- 10$^{5}$
yrs.  In this scenario, a diffuse radio halo filled with old electrons
from previous periods of activity should be apparent around CSS
sources like 3C~186 in deep, low-frequency radio maps. Most CSS and
GPS sources do not show very large-scale (100's kpc) extended radio
emission in present data (O'Dea 1998 and references therein). X-ray
observations of nearby clusters suggest a longer period of the
intermittency, e.g. 10$^6-10^7$~years (e.g. Fabian et al 2003, Forman
et al 2004). For these longer timescales 3C~186 would still be at an
early phase of the cycle.

 If there was a previous outburst we might be able to detect
``bubbles'' filled with the old radio plasma (seen as depressions in
the X-ray surface brightness) in deeper X-ray observations in the
future. The ``bubbles'' observed in nearby clusters indicate locations
of radio plasma injected into the cluster in the past (Churazov et al
2001). They are buoyant, moving within the cluster gas and last for at
least one cycle, as indicated by more than one pair of ``bubbles''
observed in some X-ray clusters (Fabian et al. 2002, Belsole et al
2001, Owen et al. 2000). 
Even if the radio source outburst lasted only $\sim$10$^5$~years the
radio plasma would still be present within buoyant ``bubbles'' loosing
its energy via synchrotron radiation.  Thus if there was a previous
outburst we might be able to detect the radio relic in low-frequency
observations.

At large redshifts, large-scale radio halos may produce a
non-thermal X-ray component via inverse Compton scattering off the
CMB. In the case of 3C~186, at z=1.063, the X-ray spectral fits do not
require such a power-law component in addition to the thermal
emission. From our data (Sec.2.2), we estimate that a non-thermal
component contributes less than 12$\%$ of the total emission,
F(1~keV)$\sim$few~nJy.

No large-scale extended radio emission is visible in our 1.5 GHz VLA
image -- we estimate that there is less than 0.3 Jy coming from a
possible halo by integrating the 3$\sigma$ rms limit over the extent
of the detected X-ray emission ($\sim$15''). The electrons scattering
CMB photons to 1 keV ($\gamma\sim$10$^{3}$) radiate synchrotron radio
emission at much lower frequencies ($\nu \sim$ (B / 1 $\mu$G ) MHz). A
deep low-frequency radio observation aimed at detecting extended radio
emission on the scale of the X-ray cluster would put a useful lower
limit on the cluster magnetic field assuming equipartition (see
Carilli \& Taylor 2002).

\section{Summary}

Our main results can be summarize as follows:

\begin{itemize}

\item[1] We have observed the CSS quasar 3C~186 for 38~ksec and detected
X-ray cluster emission extending out to $\sim 120$~kpc from the quasar.

\item[2] The cluster temperature and luminosity follows the relationship 
observed in the other high redshift clusters, its gas mass is
relatively small. The low mass could be the result of a small core
radius measured in this cluster in comparison to the other clusters at
$z>0.7$.

\item[3] The estimated pressure of the cluster gas is 2-3 order of magnitude
lower than the pressure of the radio components, thus the cluster gas
cannot confine the expanding radio source.

\item[4] Non-thermal and thermal emission associated with the radio 
components cannot be separated spatialy from the unresolved X-ray
emission that is measured. We have placed an upper limit of 1.5$\times
10^{44}$~erg~sec$^{-1}$ on the contribution from the gas that is shock
heated as a result of the radio-source expansion.  Detailed modeling
is required to estimate the relative contribution of several possible
components identified in radio observations.

\item[5] Future low frequency radio observation may provide information 
about the large scale distribution of an old electron population from
a previous outburst of the quasar activity in this source.

\end{itemize}

\acknowledgments

We thank Paul Nulsen and Brian McNamara for useful discussions. We
thank the anonymous referee for a careful reading of the manuscript and
comments. This research is funded in part by NASA contract
NAS8-39073. Partial support for this work was provided by the National
Aeronautics and Space Administration through Chandra Award Number
GO-01164X and GO2-3148A issued by the Chandra X-Ray Observatory
Center, which is operated by the Smithsonian Astrophysical Observatory
for and on behalf of NASA under contract NAS8-39073.  The VLA is a
facility of the National Radio Astronomy Observatory is operated by
Associated Universities, Inc. under a cooperative agreement with the
National Science Foundation.

This work was supported in part by NASA grants GO2-3148A,
GO-09820.01-A and NAS8-39073.



\begin{table}
\begin{scriptsize}
\caption{Spectral Models of the X-ray Diffuse Emission}
\medskip
\begin{tabular}{lccccccccccl}
\hline
\hline\noalign{\smallskip}
Model$^a$ & N$_H$ or $z$  & Abund & $\Gamma$ & Temp & Norm$^b$  
&  $\chi{2}$$^c$\\
 &   &  Solar &  & keV & $10^{-5}$ & 
(DOF)\\
\hline\hline
\\
PL & 9.15$^{+5.12}_{-3.89}$ & - & 2.06$^{+0.27}_{-0.22}$ & - & 3.08$^{+0.66}_{-0.48}$ & 
79.92 (73) \\
\\
PL+Gaussian Line &  & - & 2.12$^{\pm0.12}$ &  & 3.06$^{\pm0.21}$ 
& 68.69 (71) \\
E$_{l}^d$ = 3.18$\pm{0.07}$\\
EW=412.7eV \\
\\
RAYMOND & -  & 1 & & 4.6$^{+1.1}_{-0.7}$ & 30.2$^{+2.3}_{-2.3}$ & 
73.7 (74) \\ 
\\
RAYMOND & -  & 0.3 & -  & 5.2$^{+1.3}_{-0.9}$ & 34.8$^{+3.1}_{-3.0}$ 
&  74.0 (74) \\
\\
RAYMOND &  1.11$^{+0.04}_{-0.03}$  & 0.3 &  & 5.4$^{+1.4}_{-1.0}$ & 36.3$^{+3.4}_{-3.5}$ 
&  71.1(74) \\
\\
\\
MEKAL &   -  & 1 &  - & 4.6$^{+1.1}_{-0.7}$ & 29.8$^{+2.2}_{-2.2}$ 
&  71.9 (74) \\
\\
MEKAL &   -   & 0.3 & -  & 5.2$^{+1.3}_{-0.9}$ & 34.6$^{+3.0}_{-3.0}$ & 
73.7 (74) \\
\\
\hline
\hline\noalign{\smallskip}
\end{tabular}

 
$^a$ All models except the first PL model include equivalent Hydrogen
column of 5.68$\times 10^{20}$~cm$^{-2}$ in the Milky Way from COLDEN;
$^b$ Normalization in 10$^{-5}$~photons~cm$^2$~sec$^{-1}$ at 1~keV for
a power law model; for thermal models the normalization is defined as
$Norm \times 10^{-14} / (4 \pi (D (1+z))^2) \int n_e n_H dV$
following the definitions of RAYMOND and MEKAL models in XSPEC,
abundance table set to Anders \& Grevesse (1989);
$^c$ $\chi^2$ following Primini et al in {\it Sherpa};
$^d$ Energy of the emission line in keV.
\end{scriptsize}
\label{tab:tab1}
\end{table}

\begin{table}
\begin{scriptsize}
\caption{Spatial Models of the X-ray Diffuse Emission}
\medskip
\label{tab2}
\begin{tabular}{lccccccccccl}
\hline
\hline\noalign{\smallskip}
Region & Counts  & Net Counts & Net Counts & Soft & Hard & Soft/Hard \\ 
 & Tot & Tot & 0.3-7keV  & 0.5-2keV & 2-10keV & \\
\hline
&\\
Entire Annulus & 1189$\pm35$ & 741.1$\pm40.5$ & 691.0$\pm32.5$ &
$562.1^{+31.1}_{-30.3}$ & $143.9^{+20.9}_{-20.0}$ & $3.9^{+0.6}_{-0.8}$ 
\\
2.7$^{\prime\prime}$-15$^{\prime\prime}$ &\\
&\\
Inner Annulus & 490$\pm22$ & 392.9$\pm24.2$ & 379.3$\pm21.4$ &
$317.2^{+20.9}_{-21.4}$ & $63.2^{+10.6}_{-11.3}$ & $5.2^{+0.9}_{-1.2}$ & 
\\
2.7$^{\prime\prime}$-7.8$^{\prime\prime}$ &\\
&\\
Outer Annulus & 699$\pm26$ & 348.2$\pm32.4$ & 311.6$\pm24.5$  &
$247.1^{+21.7}_{-21.8}$ & $82.8^{+16.9}_{-16.3}$ & $3.1^{+0.4}_{-0.6}$ 
&\\
7.8$^{\prime\prime}$-15$^{\prime\prime}$ &\\
&\\
Sector 1& 206$\pm14$ & 142.1$\pm16.4$ & 128.1$\pm13.4$ 
& $100.4^{+15.5}_{-16.0}$  & $37.6^{+8.1}_{-8.9}$ &
$2.8^{+0.7}_{-0.9}$ &
&\\
&\\
Sector 2& 430$\pm21$ & 272.2$\pm24.2$ & 246.1$\pm19.3$ &
$200.9^{+22.9}_{-23.1}$ & $60.6^{+11.7}_{-12.2}$ & $3.4^{+0.5}_{-0.7}$ & 
\\
&\\
Sector 3& 164$\pm13$ & 111.2$\pm14.7$ & 107.6$\pm12.3$ & 
$96.7^{+15.2}_{-15.6}$ 
& $11.9^{+5.5}_{-6.2}$ & $10.4^{+1.1}_{-6.1}$ & 
\\
&\\
Sector 4& 395$\pm20$ & 230.8$\pm23.6$ & 219.6$\pm18.8$ &
$181.4^{+16.8}_{-17.5}$ & $34.7^{+10.5}_{-10.6}$ & $5.7^{+1.2}_{-2.2}$ & 
\\
&\\
\hline
\hline\noalign{\smallskip}
\end{tabular}


\end{scriptsize}
\label{tab:tab2}
\end{table}

\begin{table}
\caption[]{\label{table-3} Summary of Spatial and Spectral Fits to Diffuse X-ray Emission}
\begin{center}
\begin{tabular}{ll}
\hline \hline
Parameter & Property \\
\hline
$\beta$-model (1D)              & $\beta$=0.64$^{+0.11}_{-0.07}$, $r_{\rm c}$=5.8$^{+2.1}_{-1.7}\arcsec$ \\
$\beta$-model (2D)              & $\beta$=0.58$^{+0.06}_{-0.05}$, $r_{\rm c}$=5.5$^{+1.5}_{-1.2}\arcsec$  \\
                                & ellipticity = 0.24$^{+0.06}_{-0.07}$, PA=47$\pm 10$ degrees\\
$E_{\rm obs}$ (Fe-line)         & 3.18$\pm$0.07 keV\\
EW (Fe-line)                    & 412 eV\\
$F_{\rm obs}$(0.5--2 keV)       & 6.2 $\pm$ 0.3 $\times$ 10$^{-14}$ erg sec$^{-1}$ cm$^{-2}$\\
$F_{\rm obs}$(2--10 keV)        & 5.0 $\pm$ 0.7 $\times$ 10$^{-14}$ ergs sec$^{-1}$ cm$^{-2}$ \\
$F_{\rm nonthermal}$ (1 keV)    & $<$5.4 $\times$ 10$^{-15}$ erg sec$^{-1}$ cm$^{-2}$\\
$L_{\rm tot}$(0.5--2 keV)       & $6 \times$ 10$^{44}$ erg sec$^{-1}$\\

\hline \hline
\end{tabular}
\end{center}

{\footnotesize

Fluxes are unabsorbed. Luminosities are K-corrected and in the source frame. (see text).}
\end{table}

\begin{table}
\begin{scriptsize}
\caption{Quasar Models}
\medskip
\begin{tabular}{lccccccccccl}
\hline
\hline\noalign{\smallskip}
Model$^a$ & N$(z_{qso})^b$ & $\Gamma$ & Flux$^c$ & E$_{line}$ & EW$_{line}$ & $\chi^{2}$\\
   &  &  & [1 keV] & [keV] & [eV] & (DOF)$^d$\\
\hline
\\
Power Law & $<9.0^e$  & 2.01$\pm0.07$ & 7.52$\pm$0.32 & & & 138.2 (137)
\\ PL+Gaussian Line && 2.03$\pm$0.07 & 7.48$\pm$0.28 &
3.07$^{^{+0.07}_{-0.16}}$ & 130 & 132.1 (135) \\
\hline
\hline\noalign{\smallskip}
\end{tabular}

$^a$ model assumes 5.68$\times 10^{20}$ atoms~cm$^{-2}$ equivalent
Hydrogen column in the Milky Way from COLDEN; \\ 
$^b$ absorbing column intrinsic to the quasar at z=1.063 in 
units of $10^{20}$ atoms~cm$^{-2}$;\\
$^c$ flux in units of 10$^{-5}$~photons~cm$^{-2}$~sec$^{-1}$~keV$^{-1}$
$^d$ $\chi^2$ following formulation of Primini et al. in {\it Sherpa};
$^e$ 3$\sigma$ upper limit.

\end{scriptsize}
\label{tab:tab3}
\end{table}


\begin{figure}
\epsscale{0.85}
\caption{Adaptively smoothed exposure corrected image (photons energies within 
0.3-7~keV range) of the {\it Chandra} ACIS-S observation of 3C~186
(Q0740+380) The diffuse emission is detected on $>100$~kpc scale,
1\arcsec=8.2~kpc. North is up and East is left. Contours represent a
surface brightness of: (0.046,0.066, 0.13, 0.2, 0.33, 0.46, 0.66,
6.635, 33.175)$\times 10^{-6}$ photons~cm$^{-2}$~arcsec$^{-2}$.
The direction of the CCD readout is indicated by arrow on the right
side.  A blue arrow in the upper right corner shows the PA=-37~deg of 
the 2$\arcsec$ radio jet (see Fig~\ref{fig:radio}.}
\label{fig:smooth}
\end{figure}


\begin{figure}
\epsscale{1.0}
\caption{Adaptively smoothed exposure corrected image of the
{\it Chandra} ACIS-S observation 3C~186 (Q0740+380) in two X-ray
bands: (1) soft 0.5-2keV on the left and (2) hard 2-7~keV on the
right. North is up and East is left. Contours represent a surface
brightness of: (0.036,0.066, 0.11, 0.3, 0.66, 6.635, 33.175)$\times
10^{-6}$ photons~cm$^{-2}$~arcsec$^{-2}$. 1\arcsec=8.2~kpc.
}
\label{fig:comparison}
\end{figure}


\begin{figure}
\epsscale{0.85}
\plotone{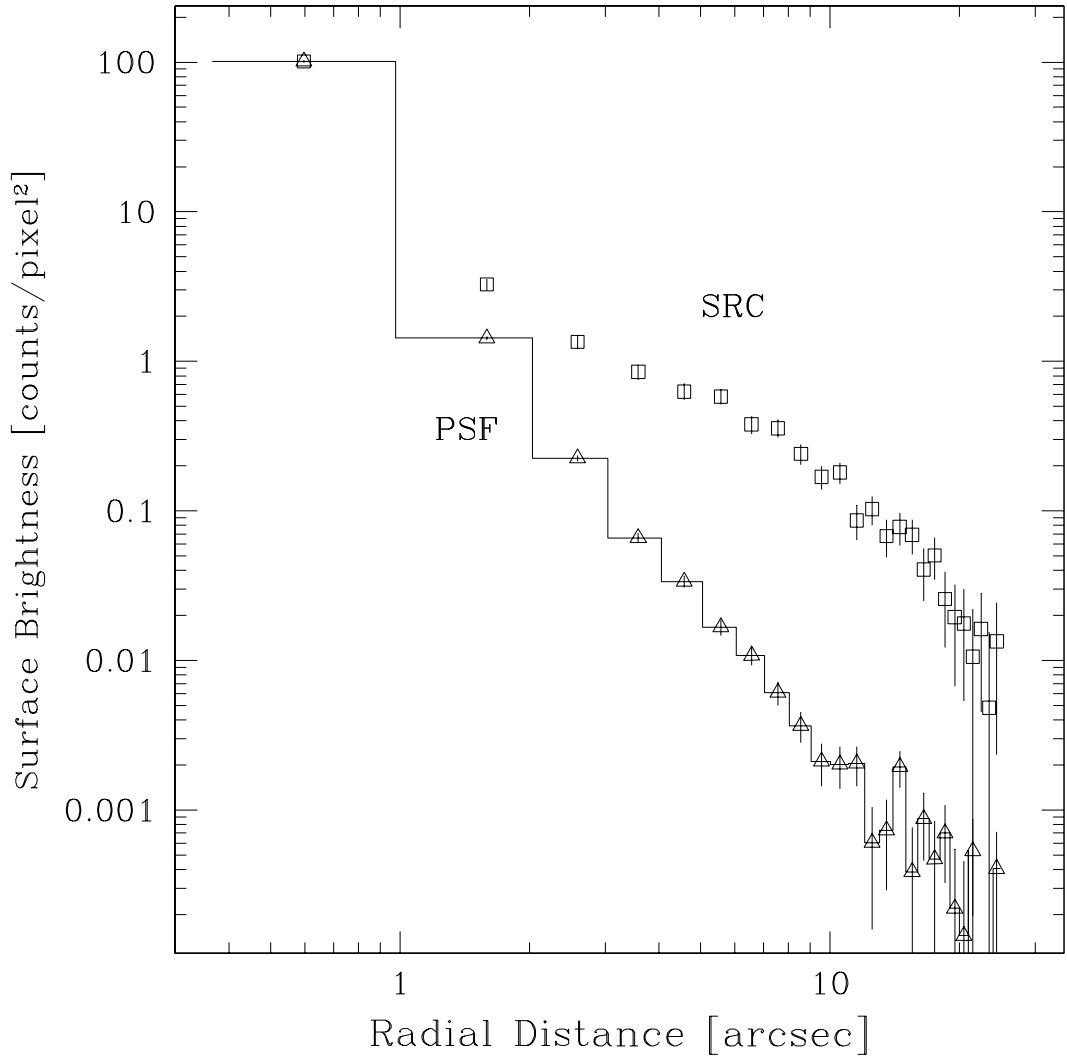}
\caption{The background subtracted 
surface brightness profile up to 25\arcsec distance from the
quasar. The source data is indicated by the square points. The solid
line with triangular points represents the simulated PSF profile. 
}
\label{fig:sim1}
\end{figure}

\begin{figure}
\epsscale{0.85}
\plotone{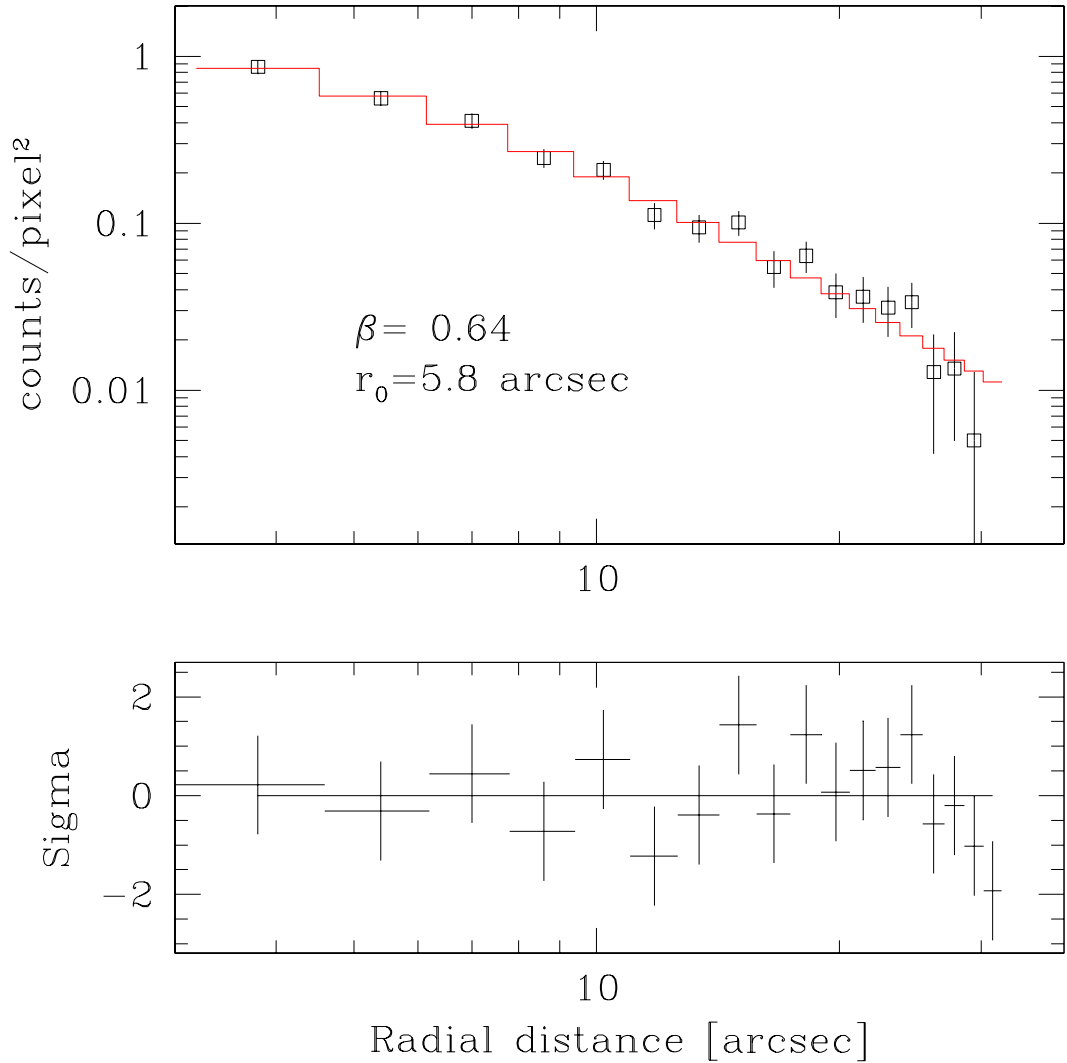}
\caption{Background subtracted surface brightness profile for
radii $3\arcsec$ to 30$\arcsec$ fit with a beta model.  The data are
indicated by the square points. The solid line shows the best fit
model with parameter $\beta=0.64^{+0.11}_{-0.07}$ and a core radius of
$r_{core}=5.8\arcsec^{+2.1}_{-1.7}$. The bottom panel illustrates the
differences between the data and the model in units of $\sigma$.}
\label{fig:sim2}
\end{figure}


\begin{figure}
\epsscale{0.85}
\plotone{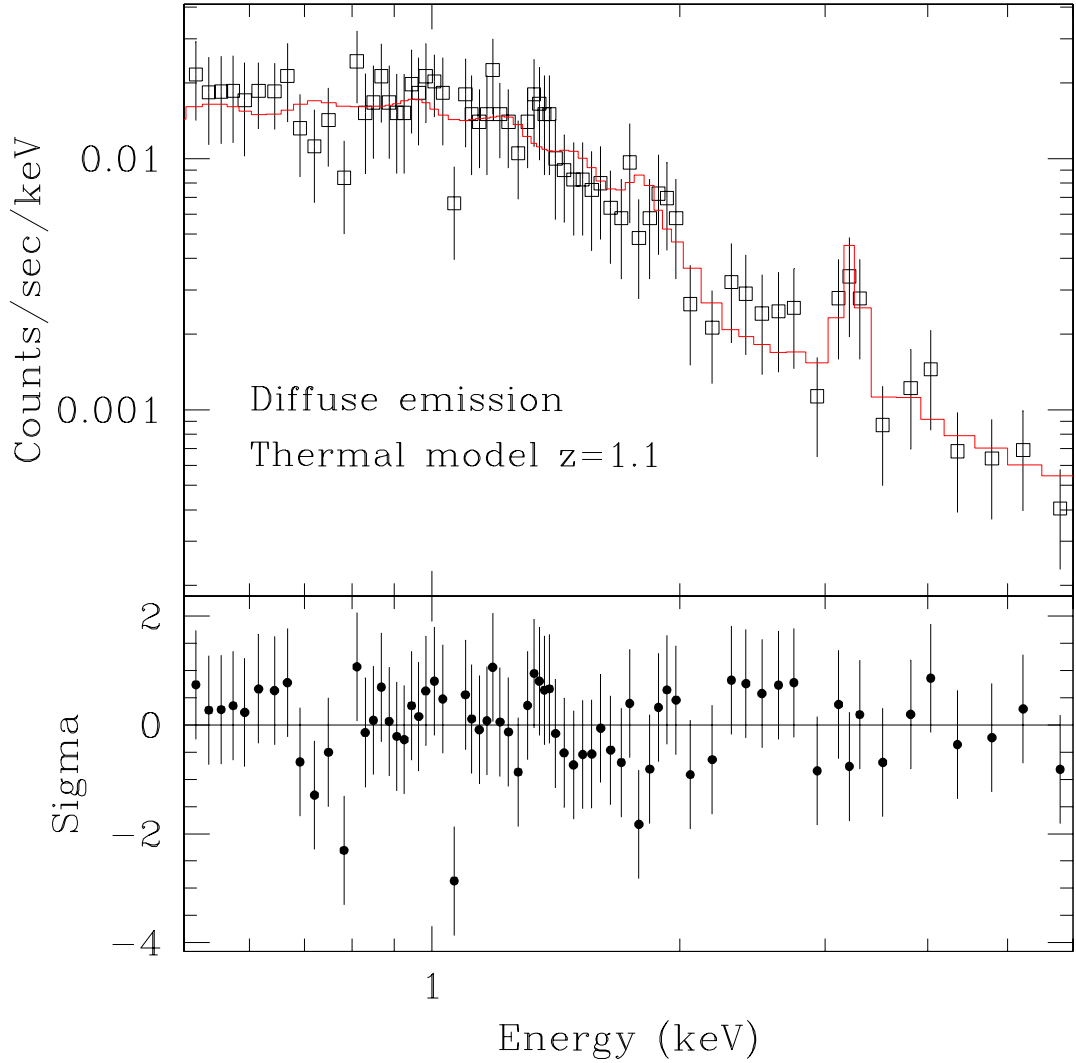}
\caption{ACIS-S spectrum of the diffuse emission between 2.7 and 15 arcsec 
which we fit with the plasma model. Upper panel shows the data
indicated by squares with 1$\sigma$ error-bars and the best-fit model
drawn with a solid line. The bottom panel shows the residual
difference between model and the data in units of sigma. The scatter
is due to calibration uncertainties.}
\label{fig:raymond}
\end{figure}


\begin{figure}
\epsscale{0.9}
\plottwo{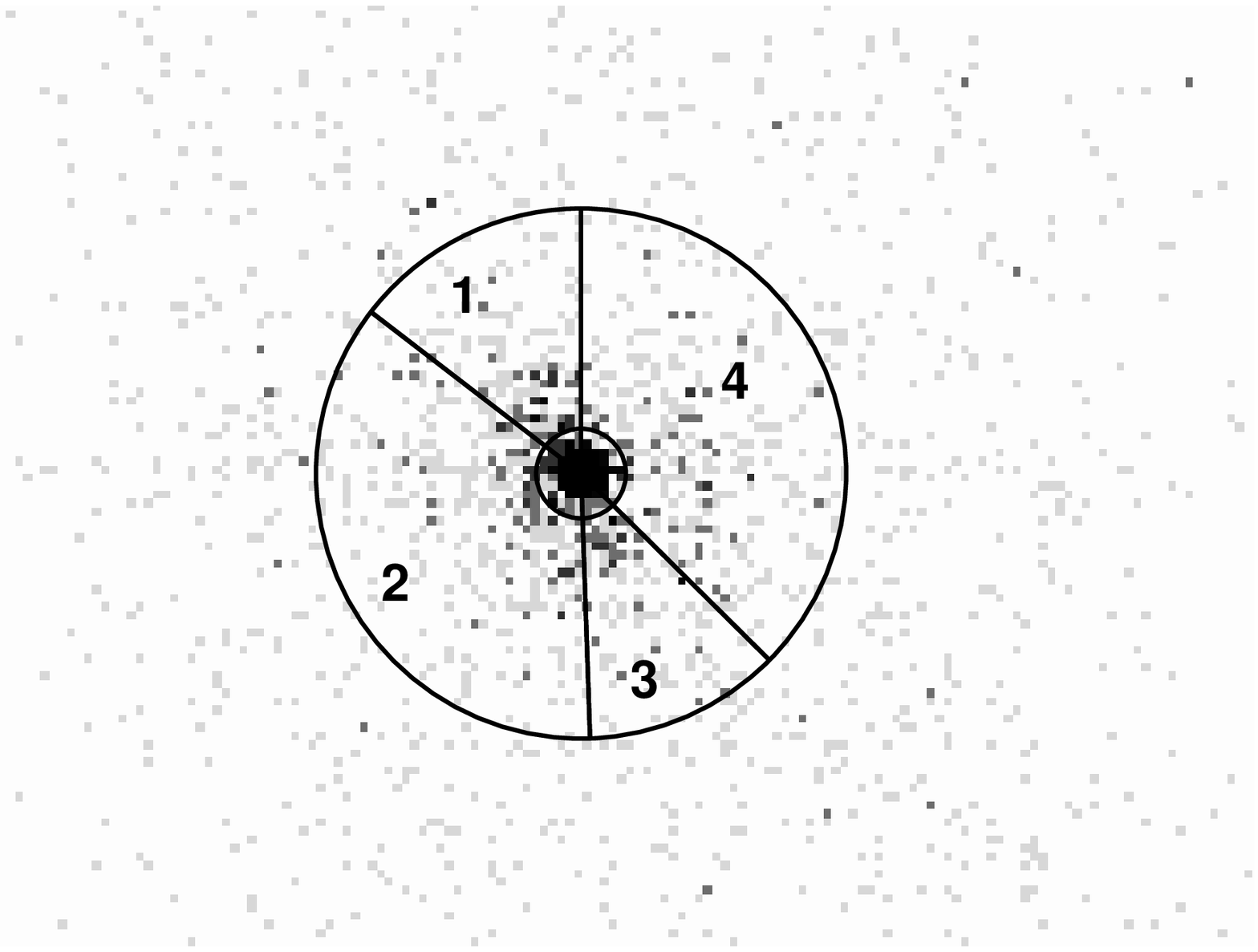}{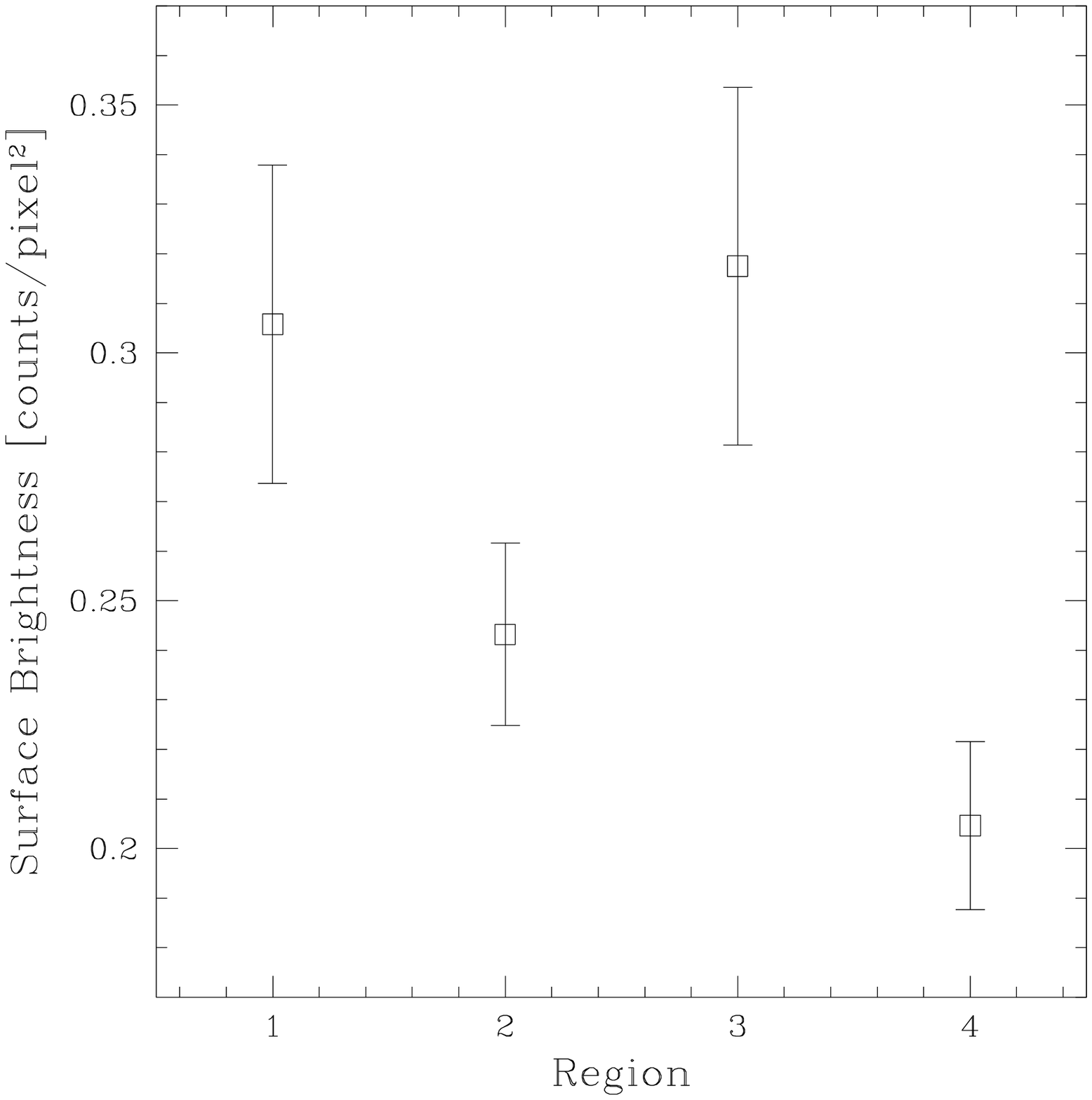}
\caption{{\bf Left:} Defined sector regions overlayed 
over an {\it Chandra} ACIS-S image of 3C~186. One pixel corresponds to
0.492\arcsec. North is up and East is left {\bf Right:} Background
subtracted surface brightness calculated for each sector in the left
panel. The sector numbers are indicated on the bottom axis.}
\label{fig:panda}
\end{figure}


\begin{figure}
\epsscale{0.9}
\plotone{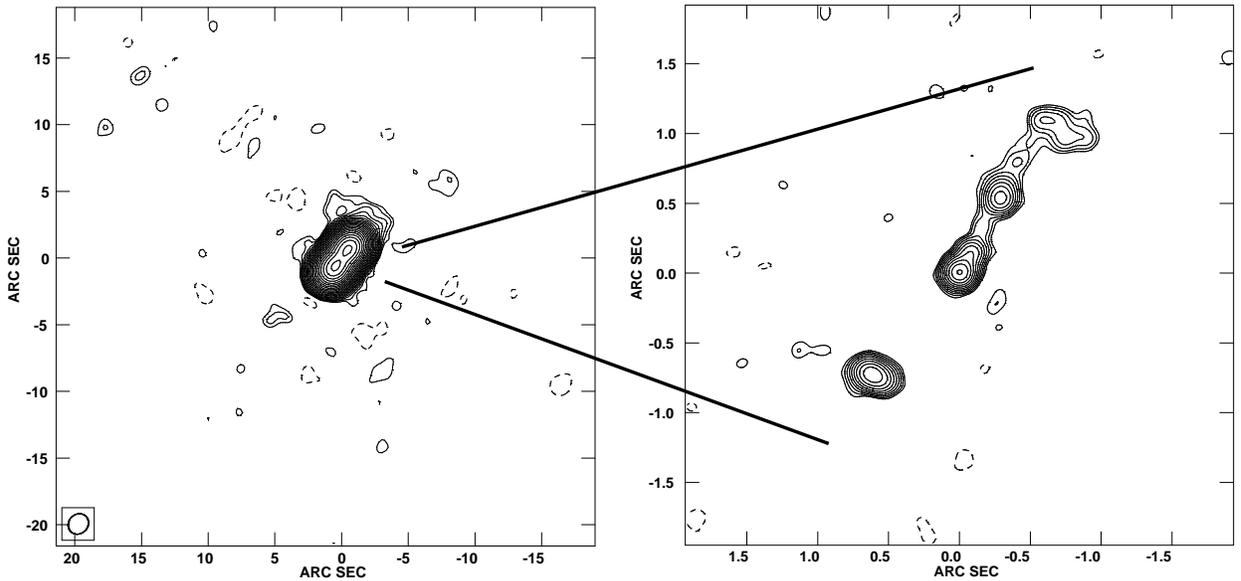}
\caption{{\bf Left:} VLA 1.5 GHz image of 3C~186 from reprocessing the A- and 
B-configuration archival datasets published in van Breugel et
al. (1992). The restoring beam is 1.62'' $\times$ 1.44'' at position
angle --42.7 degrees shown at bottom left. The image peak is
565~mJy/bm and contour levels begin at 0.5 mJy/beam (2$\sigma$) and
increase by factors of $\sqrt{2}$. Some extended radio emission is
apparent, though not at the angular scale of the observed extended
X-rays. North is up East is left.  {\bf Right:} High resolution
(0.15$\arcsec$) VLA 15~GHz image of 3C~186 showing the core-jet
morphology of the source. The image peak is 21.6 mJy/beam, and
contours begin at 0.65 mJy/beam increasing by factors of $\sqrt{2}$.}
\label{fig:radio}
\end{figure}

\begin{figure}
\epsscale{0.9}
\plottwo{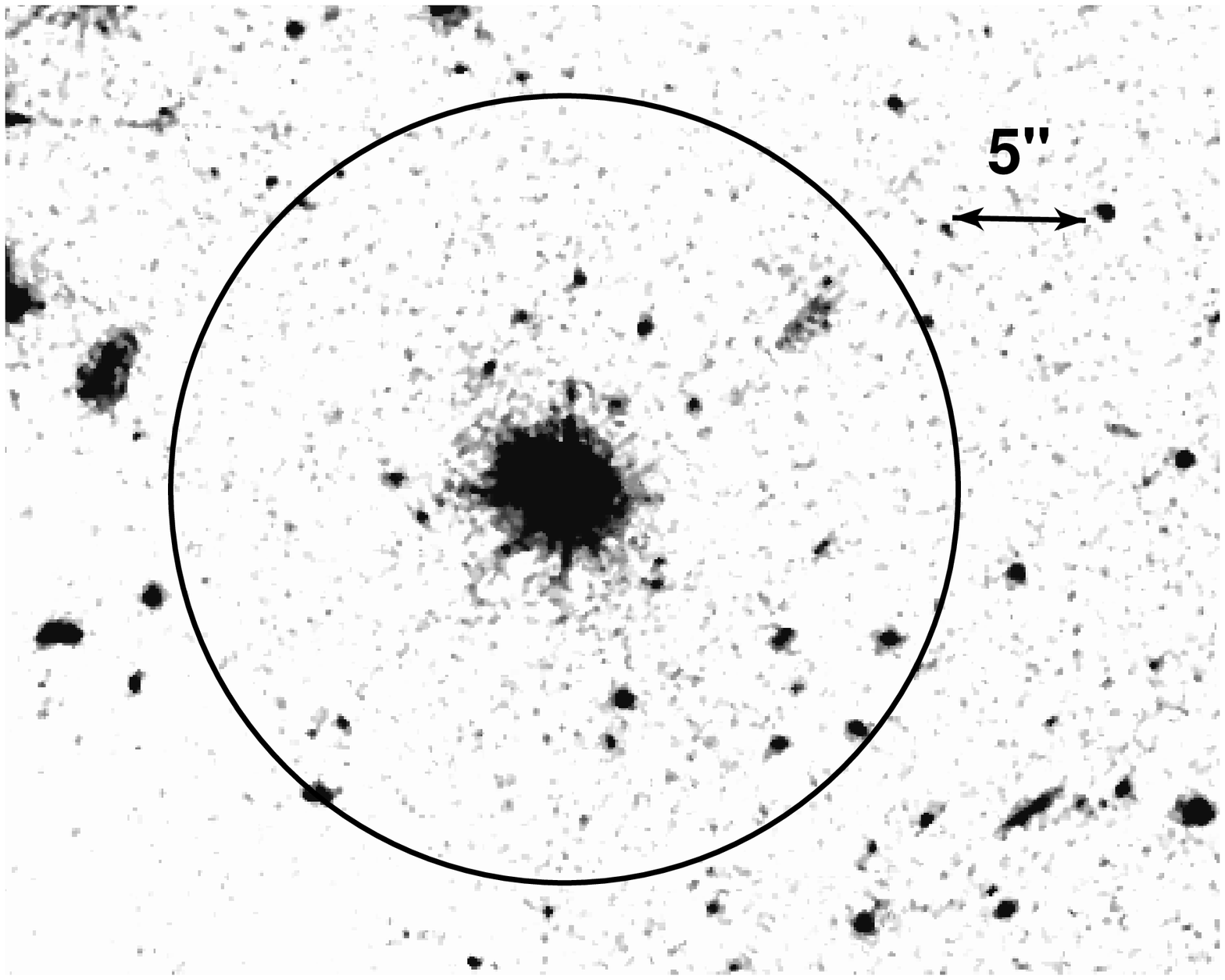}{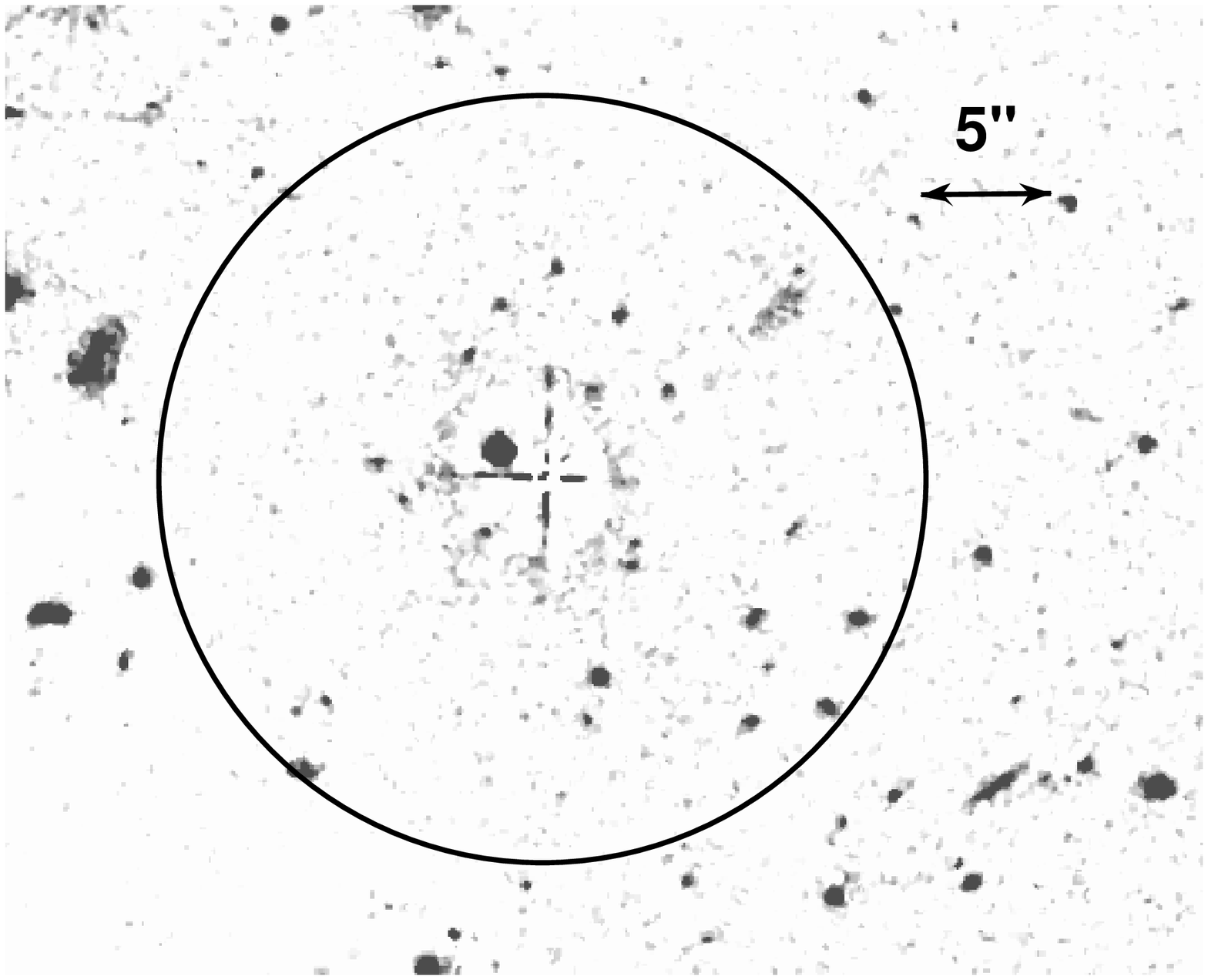}
\caption{ {\bf Left:} HST image of the field. The 15\arcsec\ radius circle highlights the region of the diffuse X-ray emission. North is up East is
left. {\bf Right:} The host galaxy contribution subtracted from the
central image. The ring of diffuse emission between 2-3$\arcsec$ is an
artifact of the imperfect galaxy subtraction. Also, the four
diffraction spikes (the cross at the center) are apparent.}
\label{fig:hst}
\end{figure}


\begin{figure}
\epsscale{0.85}
\plotone{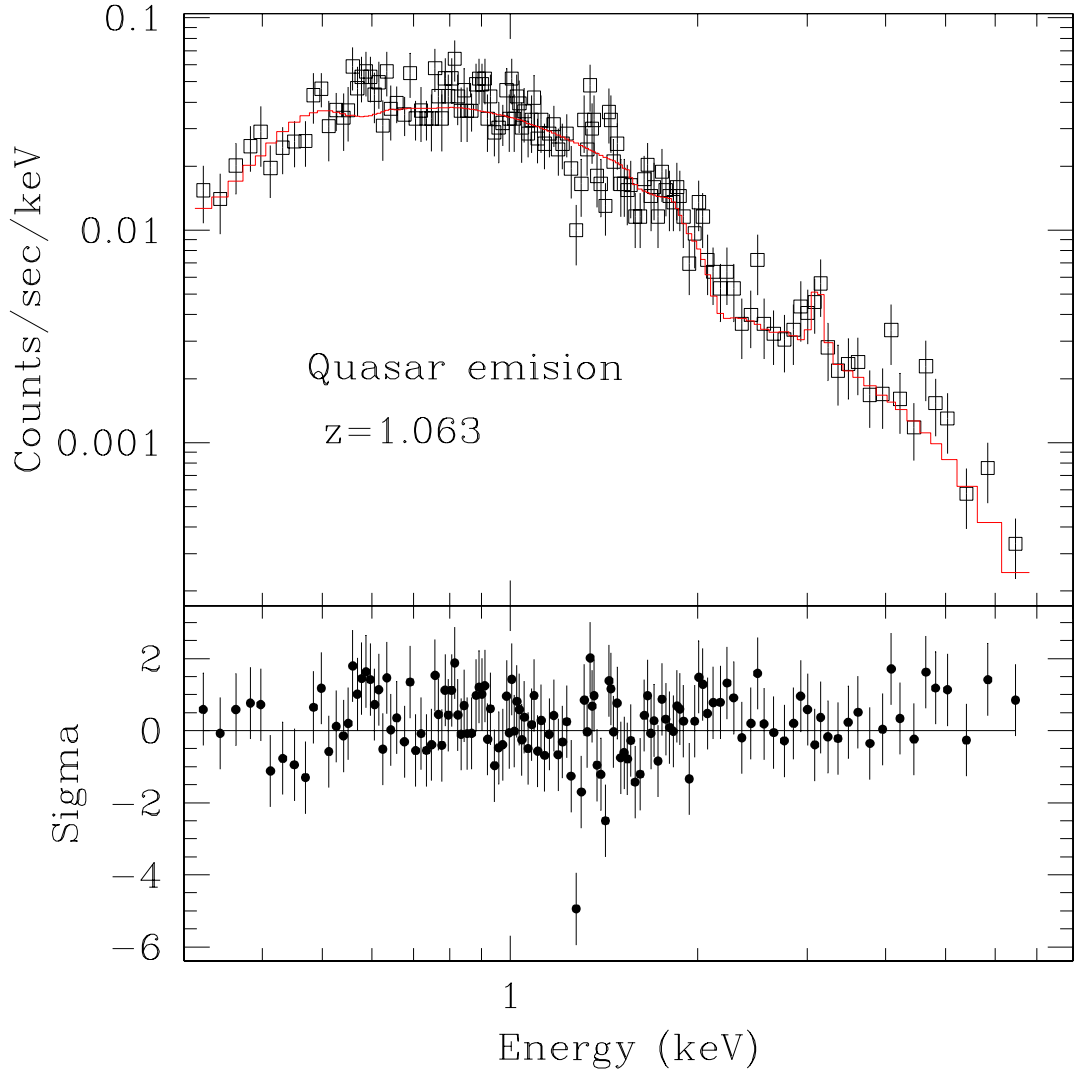}
\caption{Upper panel shows the ACIS-S spectrum of the z=1.063 quasar 
Q0740+380 (empty squares) over-plotted with the best-fit power law
plus Gaussian line model at E$_{obs}=3.07^{+0.06}_{-0.11}$~keV (a
solid line). The bottom panel shows the difference between model and
the data in units of sigma. The scatter between 1~keV and 2~keV is due
to calibration uncertainties.}
\label{fig:qsofit}
\end{figure}


\begin{thebibliography}{}
\setlength\itemsep{0cm}

\bibitem[Alexander(2000)]{2000MNRAS.319....8A} Alexander, P.\ 2000, \mnras, 
319, 8 

\bibitem[Anders \& Grevesse (1989)]{anders}
Anders E. \& Grevesse N.\ 1989, Geochimica et Cosmochimica Acta 53, 197 

\bibitem[Barr et al.(2003)]{2003MNRAS.346..229B} Barr, J.~M., Bremer, 
M.~N., Baker, J.~C., \& Lehnert, M.~D.\ 2003, \mnras, 346, 229 
								
\bibitem[Baum, O'Dea, de Bruyn, \& Murphy(1990)]{baum} Baum, 
S.~A., O'Dea, C.~P., de Bruyn, A.~G., \& Murphy, D.~W.\ 1990, A\& A, 232, 19 

\bibitem[Bechtold et al.(1994)]{1994AJ....108..759B} Bechtold, J., et al.\ 
1994, \aj, 108, 759 

\bibitem[Belsole et al.(2001)]{2001A&A...365L.188B} Belsole, E., et al.\ 
2001, \aap, 365, L188 

\bibitem[Bicknell et al..(1997)]{bicknell} Bicknell, G.~V., 
Dopita, M.~A., \& O'Dea, C.~P.\ 1997, \apj, 485, 112 


\bibitem[Bremer et al.(1997)]{1997MNRAS.284..213B} Bremer, M.~N., Fabian, 
A.~C., \& Crawford, C.~S.\ 1997, \mnras, 284, 213 

\bibitem[Cash (1979)]{cash79} Cash, W.\ 1979, \apj, 228, 939

\bibitem[Carilli et al..(2002)]{2002ApJ...567..781C} Carilli, C.~L., Harris, 
D.~E., Pentericci, L., R{\" o}ttgering, H.~J.~A., Miley, G.~K., Kurk, 
J.~D., \& van Breugel, W.\ 2002, \apj, 567, 781 

\bibitem[Carilli \& Taylor(2002)]{2002ARA&A..40..319C} Carilli, C.~L., \& 
Taylor, G.~B.\ 2002, \araa, 40, 319 

\bibitem[Carvalho(1985)]{1985MNRAS.215..463C} Carvalho, J.~C.\ 1985, 
\mnras, 215, 463

\bibitem[Carvalho(1998)]{1998A&A...329..845C} Carvalho, J.~C.\ 1998, \aap, 
329, 845 

\bibitem[Cawthorne et al..(1986)]{1986MNRAS.219..883C} Cawthorne, T.~V.,
Scheuer, P.~A.~G., Morison, I., \& Muxlow, T.~W.~B.\ 1986, \mnras, 219,883

\bibitem[Celotti \& Fabian(2004)]{2004MNRAS.tmp..261C} Celotti, A.~\& 
Fabian, A.~C.\ 2004, \mnras, 261 

\bibitem[Crawford \& Fabian(2003)]{2003MNRAS.339.1163C} Crawford, C.~S.~\& 
Fabian, A.~C.\ 2003, \mnras, 339, 1163 

\bibitem[Churazov et al.(2001)]{2001ApJ...554..261C} Churazov, E., Br{\" 
u}ggen, M., Kaiser, C.~R., B{\" o}hringer, H., \& Forman, W.\ 2001, \apj, 
554, 261 



\bibitem[De Young(1991)]{1991ApJ...371...69D} De Young, D.~S.\ 1991, \apj, 
371, 69 
 
\bibitem[Di Matteo et al.(2005)]{tizi} Di Matteo, T., 
Springel, V., \& Hernquist, L.\ 2005, \nat, 433, 604 

\bibitem[Donahue et al..(2003)]{2003ApJ...598..190D} Donahue, M., Gaskin, 
J.~A., Patel, S.~K., Joy, M., Clowe, D., \& Hughes, J.~P.\ 2003, \apj, 598, 
190 

\bibitem[Elvis et al..(1994)]{elvis} Elvis, M., Fiore, F., 
Wilkes, B., McDowell, J., \& Bechtold, J.\ 1994, ApJ, 422, 60 

\bibitem[Elvis et al.(1985)]{1985ApJ...292..357E} Elvis, M., Wilkes, B.~J., 
\& Tananbaum, H.\ 1985, \apj, 292, 357 

\bibitem[Ellingson, Yee, \& Green(1991)]{ellingson2} Ellingson, 
E., Yee, H.~K.~C., \& Green, R.~F.\ 1991, \apj, 371, 49 

\bibitem[Evans \& Koratkar(2004)]{2004ApJS..150...73E} Evans, I.~N., \& 
Koratkar, A.~P.\ 2004, \apjs, 150, 73 

\bibitem[Yee \& Ellingson(1993)]{ellingson3} Yee, H.~K.~C.~\& 
Ellingson, E.\ 1993, \apj, 411, 43 

\bibitem[Fabian et al.(2005)]{2005astro.ph..1222F} Fabian, A.~C.,
Reynolds, C.~S., Taylor, G.~B., \& Dunn, R.~J.~H.\ 2005, MNRAS, submitted
(astro-ph/0501222)

\bibitem[Fabian(1994)]{1994ARA&A..32..277F} Fabian, A.~C.\ 1994, \araa, 32, 
277 

\bibitem[Fabian et al..(2003)]{2003MNRAS.341..729F} Fabian, A.~C., Sanders, 
J.~S., Crawford, C.~S., \& Ettori, S.\ 2003, \mnras, 341, 729 

\bibitem[Fabian et al..(2003)]{2003MNRAS.344L..43F} Fabian, A.~C., Sanders, 
J.~S., Allen, S.~W., Crawford, C.~S., Iwasawa, K., Johnstone, R.~M., 
Schmidt, R.~W., \& Taylor, G.~B.\ 2003, \mnras, 344, L43 

\bibitem[Fabian et al.(2002)]{2002MNRAS.331..369F} Fabian, A.~C., Celotti, 
A., Blundell, K.~M., Kassim, N.~E., \& Perley, R.~A.\ 2002, \mnras, 331, 
369 

\bibitem[Fabian \& Nulsen(1977)]{1977MNRAS.180..479F} Fabian, A.~C., \& 
Nulsen, P.~E.~J.\ 1977, \mnras, 180, 479

\bibitem[Fanti et al..(1995)]{1995A&A...302..317F} Fanti, C., Fanti, R., 
Dallacasa, D., Schilizzi, R.~T., Spencer, R.~E., \& Stanghellini, C.\ 1995, 
\aap, 302, 317 


\bibitem[Forman et al. (2003)]{m87..outburst} Forman, W. et al astro-ph/0312576

\bibitem[Freeman et al..(2001)]{2001SPIE.4477...76F} Freeman, P., Doe, S., 
\& Siemiginowska, A.\ 2001, \procspie, 4477, 76 

\bibitem[Guainazzi et al.(2004)]{matteo} Guainazzi, M., 
Siemiginowska, A., Rodriguez-Pascual, P., \& Stanghellini, C.\ 2004, \aap, 
421, 461 

\bibitem[Gugliucci et al.(2005)]{2005ApJ...622..136G} Gugliucci, N.~E., 
Taylor, G.~B., Peck, A.~B., \& Giroletti, M.\ 2005, \apj, 622, 136 

\bibitem[Hales et al.(1993)]{1993MNRAS.263...25H} Hales, S.~E.~G., Baldwin, 
J.~E., \& Warner, P.~J.\ 1993, \mnras, 263, 25

\bibitem[Hardcastle \& Worrall(1999)]{1999MNRAS.309..969H} Hardcastle, 
M.~J., \& Worrall, D.~M.\ 1999, \mnras, 309, 969 

\bibitem[Heinz et al.(1998)]{1998ApJ...501..126H} Heinz, S., Reynolds, 
C.~S., \& Begelman, M.~C.\ 1998, \apj, 501, 126 

\bibitem[Janiuk et al.(2004)]{2004ApJ...602..595J} Janiuk, A., Czerny, B., 
Siemiginowska, A., \& Szczerba, R.\ 2004, \apj, 602, 595 

\bibitem[Jerius et al.(2004)]{2004SPIE.5165..433J} Jerius, D.~H., Gaetz, 
T.~J., \& Karovska, M.\ 2004, \procspie, 5165, 433 

\bibitem[Jeyakumar et al.(2005)]{2005A&A...432..823J} Jeyakumar, S., Wiita, 
P.~J., Saikia, D.~J., \& Hooda, J.~S.\ 2005, \aap, 432, 823 


\bibitem[Kuraszkiewicz et al..(2002)]{2002ApJS..143..257K} Kuraszkiewicz, 
J.~K., Green, P.~J., Forster, K., Aldcroft, T.~L., Evans, I.~N., \& 
Koratkar, A.\ 2002, \apjs, 143, 257 

\bibitem[Li \& Jin(1996)]{radiopos} Li, J.~\& Jin, W.\ 1996, 
\aaps, 120, 201 

\bibitem[Lister(2003)]{lister} Lister, M.~L.\ 2003, ASP 
Conf.~Ser.~300: Radio Astronomy at the Fringe, 71 

\bibitem[Lumb et al.(2004)]{2004A&A...420..853L} Lumb, D.~H., et al.\
2004, \aap, 420, 853

\bibitem[Marshall et al.(2005)]{2005ApJS..156...13M} Marshall, H.~L., et
al.\ 2005, \apjs, 156, 13

\bibitem[Murgia et al..(1999)]{murgia} Murgia, M., Fanti, C., 
Fanti, R., Gregorini, L., Klein, U., Mack, K.-H., \& Vigotti, M.\ 1999, 
\aap, 345, 769 

\bibitem[McNamara et al. 2005]{} McNamara B. R.; Nulsen, P. E. J.; Wise, M. W.; Rafferty, D. A.; Carilli, C.; Sarazin, C. L.; Blanton, E. L. 2005, Nature, astro-ph/0411553 

\bibitem[Netzer et al..(1996)]{1996MNRAS.279..429N} Netzer, H., et al..\ 
1996, \mnras, 279, 429 

\bibitem[Nulsen et al. 2004] {} Nulsen, P. E. J.; McNamara, B. R.; Wise, M. W.; David, L. P., 2004, astro-ph/0408315

\bibitem[Nulsen et al.(2002)]{2002ApJ...568..163N} Nulsen, P.~E.~J., David, 
L.~P., McNamara, B.~R., Jones, C., Forman, W.~R., \& Wise, M.\ 2002, \apj, 
568, 163 

\bibitem[O'Dea et al..(2000)]{odea2000} O'Dea, C.~P., De Vries, 
W.~H., Worrall, D.~M., Baum, S.~A., \& Koekemoer, A.\ 2000, AJ, 119,
478

\bibitem[O'Dea(1998)]{odea1998} O'Dea, C.~P.\ 1998, PASP, 110, 
493 

\bibitem[O'Dea \& Baum(1997)]{1997AJ....113..148O} O'Dea, C.~P., \& Baum, 
S.~A.\ 1997, \aj, 113, 148 

\bibitem[O'Dea et al.(1991)]{1991ApJ...380...66O} O'Dea, C.~P., Baum, 
S.~A., \& Stanghellini, C.\ 1991, \apj, 380, 66 

\bibitem[Owen et al.(2000)]{2000ApJ...543..611O} Owen, F.~N., Eilek, J.~A., 
\& Kassim, N.~E.\ 2000, \apj, 543, 611

\bibitem[Owsianik et al..(1998)]{1998A&A...336L..37O} Owsianik, I., Conway, 
J.~E., \& Polatidis, A.~G.\ 1998, \aap, 336, L37 

\bibitem[Phillips \& Mutel(1982)]{1982A&A...106...21P} Phillips, R.~B., \& 
Mutel, R.~L.\ 1982, \aap, 106, 21 

\bibitem[Polatidis \& Conway(2003)]{2003PASA...20...69P} Polatidis, A.~G., 
\& Conway, J.~E.\ 2003, Publications of the Astronomical Society of 
Australia, 20, 69 

\bibitem[Readhead \& Hewish(1976)]{1976MNRAS.176..571R} Readhead, A.~C.~S., 
\& Hewish, A.\ 1976, \mnras, 176, 571 

\bibitem[Readhead et al.(1996)]{1996ApJ...460..612R} Readhead, A.~C.~S., 
Taylor, G.~B., Xu, W., Pearson, T.~J., Wilkinson, P.~N., \& Polatidis, 
A.~G.\ 1996, \apj, 460, 612 

\bibitem[Readhead et al.(1996)]{1996ApJ...460..634R} Readhead, A.~C.~S., 
Taylor, G.~B., Pearson, T.~J., \& Wilkinson, P.~N.\ 1996, \apj, 460, 634 
  
\bibitem[Reynolds \& Begelman(1997)]{1997ApJ...487L.135R} Reynolds, C.~S., 
\& Begelman, M.~C.\ 1997, \apjl, 487, L135 

\bibitem[Reynolds et al.(2005)]{2005MNRAS.357..242R} Reynolds, C.~S., 
McKernan, B., Fabian, A.~C., Stone, J.~M., \& Vernaleo, J.~C.\ 2005, 
\mnras, 357, 242 

\bibitem[Roychowdhury et al.(2004)]{2004ApJ...615..681R} Roychowdhury, S., 
Ruszkowski, M., Nath, B.~B., \& Begelman, M.~C.\ 2004, \apj, 615, 681  

\bibitem[Ruszkowski et al.(2004)]{2004ApJ...611..158R} Ruszkowski, M., 
Br{\" u}ggen, M., \& Begelman, M.~C.\ 2004, \apj, 611, 158 


\bibitem[Sambruna et al.(2004)]{sam} Sambruna, R.~M.,
Gambill, J.~K., Maraschi, L., Tavecchio, F., Cerutti, R., Cheung, C.~C.,
Urry, C.~M., \& Chartas, G.\ 2004, \apj, 608, 698

\bibitem[S{\' a}nchez \& Gonz{\' a}lez-Serrano(2002)]{2002A&A...396..773S}
S{\' a}nchez, S.~F., \& Gonz{\' a}lez-Serrano, J.~I.\ 2002, \aap, 396, 773

\bibitem[Schoenmakers et al..(1999)]{scho} Schoenmakers, A.~P., 
de Bruyn, A.~G., R{\"o}ttgering, H.~J.~A., \& van der Laan, H.\ 1999,
A\&A, 341, 44

\bibitem[Schwartz et al..(2000)]{2000ApJ...540L..69S} Schwartz, D.~A., et 
al.\ 2000, \apjl, 540, L69

\bibitem[Schwartz et al.(2000)]{2000SPIE.4012...28S} Schwartz, D.~A., et 
al.\ 2000a, \procspie, 4012, 28

\bibitem[Silk and Rees (1998)] {Silk} 
Silk, J., \& Rees, M.J., 1998, A\&A, 331, L1; 

\bibitem[Siemiginowska et al.(1996)]{1996ApJ...458..491S} Siemiginowska, 
A., Czerny, B., \& Kostyunin, V.\ 1996, \apj, 458, 491 

\bibitem[Siemiginowska et al..(2003a)]{siem2003a} Siemiginowska, 
A., et al..\ 2003, ApJ, 595, 643 



\bibitem[Siemiginowska et al..(2002)]{siem2002} Siemiginowska, 
A., Bechtold, J., Aldcroft, T.~L., Elvis, M., Harris, D.~E., \& Dobrzycki, 
A.\ 2002, ApJ, 570, 543 

\bibitem[Simpson \& Rawlings(2000)]{2000MNRAS.317.1023S} Simpson, C., \& 
Rawlings, S.\ 2000, \mnras, 317, 1023 

\bibitem[Shepherd, Pearson, \& Taylor(1994)]{she94} Shepherd, M.~C.,
Pearson, T.~J., \& Taylor, G.~B. 1994, BAAS, 26, 987

\bibitem[Snellen et al..(2000)]{snellen} Snellen, I.~A.~G., 
Schilizzi, R.~T., Miley, G.~K., de Bruyn, A.~G., Bremer, M.~N., \& R{\" 
o}ttgering, H.~J.~A.\ 2000, MNRAS, 319, 445 
 
\bibitem[Spencer et al..(1991)]{1991MNRAS.250..225S} Spencer, R.~E., etal.\ 
1991, \mnras, 250, 225

\bibitem[Spergel et al.(2003)]{2003ApJS..148..175S} Spergel, D.~N., et
al.\ 2003, \apjs, 148, 175

\bibitem[Stanghellini et al..(2001)]{stan2001} Stanghellini, C., 
Dallacasa, D., O'Dea, C.~P., Baum, S.~A., Fanti, R., \& Fanti, C.\
2001, A\&A, 377, 377

\bibitem[Stanghellini et al..(1998)]{stan1998} Stanghellini, C., 
O'Dea, C.~P., Dallacasa, D., Baum, S.~A., Fanti, R., \& Fanti, C.\ 1998, 
A\&AS, 131, 303 

\bibitem[Stanghellini, Baum, O'Dea, \& Morris(1990)]{stan1990} 
Stanghellini, C., Baum, S.~A., O'Dea, C.~P., \& Morris, G.~B.\ 1990, A\&A, 
233, 379 

\bibitem[Stark et al.(1992)]{1992ApJS...79...77S} Stark, A.~A., Gammie,
C.~F., Wilson, R.~W., Bally, J., Linke, R.~A., Heiles, C., \& Hurwitz,
M.\
1992, \apjs, 79, 77


\bibitem[van Breugel et al..(1992)]{1992A&A...256...56V} van Breugel,W.~J.~M., 
Fanti, C., Fanti, R., Stanghellini, C., Schilizzi, R.~T., \&Spencer,
R.~E.\ 1992, \aap, 256, 56

\bibitem[van Breugel et al.(1984)]{1984AJ.....89....5V} van Breugel, W., 
Miley, G., \& Heckman, T.\ 1984, \aj, 89, 5 

\bibitem[Vestergaard(2002)]{2002ApJ...571..733V} Vestergaard, M.\ 2002, 
\apj, 571, 733

\bibitem[Vikhlinin et al..(2002)]{2002ApJ...578L.107V} Vikhlinin, A., 
VanSpeybroeck, L., Markevitch, M., Forman, W.~R., \& Grego, L.\ 2002, 
\apjl, 578, L107 

\bibitem[Voit et al.(2003)]{2003ApJ...593..272V} Voit, G.~M., Balogh, 
M.~L., Bower, R.~G., Lacey, C.~G., \& Bryan, G.~L.\ 2003, \apj, 593, 272 
 
\bibitem[Weisskopf et al.(2003)]{2003ExA....16....1W} Weisskopf, M.~C., et 
al.\ 2003, Experimental Astronomy, 16, 1 

\bibitem[Weisskopf et al..(2002)]{2002PASP..114....1W} Weisskopf, M.~C., 
Brinkman, B., Canizares, C., Garmire, G., Murray, S., \& Van Speybroeck, 
L.~P.\ 2002, \pasp, 114, 1 

\bibitem[Wilkinson et al.(1984)]{1984Natur.308..619W} Wilkinson, P.~N., 
Booth, R.~S., Cornwell, T.~J., \& Clark, R.~R.\ 1984, \nat, 308, 619 

\bibitem[Willott et al.(1999)]{1999MNRAS.309.1017W} Willott, C.~J., 
Rawlings, S., Blundell, K.~M., \& Lacy, M.\ 1999, \mnras, 309, 1017 

\bibitem[Worrall et al..(2001)]{2001MNRAS.326.1127W} Worrall, D.~M., 
Birkinshaw, M., Hardcastle, M.~J., \& Lawrence, C.~R.\ 2001, \mnras, 326, 
1127 

\bibitem[Worrall \& Birkinshaw(2003)]{2003MNRAS.340.1261W} Worrall, D.~M., 
\& Birkinshaw, M.\ 2003, \mnras, 340, 1261 

\bibitem[Worrall \& Birkinshaw (2004)]{}Worrall, D.~M. \& Birkinshaw M., 2004, 
to appear as a book chapter in `Physics of Active Galactic Nuclei at
all Scales', eds. D. Alloin, R. Johnson, P. Lira, (Springer Verlag),
Lecture Notes in Physics series. astro-ph/0410297

\bibitem[Worrall, Hardcastle, Pearson, \& 
Readhead(2004)]{diana2004} Worrall, D.~M., Hardcastle, M.~J., Pearson,
T.~J., \& Readhead, A.~C.~S.\ 2004, MNRAS, 347, 632

\end{thebibliography}
\end{document}